\documentclass[11pt]{article}
\usepackage[T1]{fontenc}
\usepackage{lmodern}
\usepackage[margin=1in]{geometry}
\usepackage{microtype,needspace}
\usepackage{amsmath,amssymb,amsthm,mathtools}
\usepackage{booktabs,longtable,array,enumitem,graphicx,xcolor}
\usepackage[ruled]{algorithm}
\usepackage{algpseudocode}
\algrenewcommand\algorithmicrequire{\textbf{Input:}}
\algrenewcommand\algorithmicensure{\textbf{Output:}}
\usepackage[numbers,sort&compress]{natbib}
\usepackage[colorlinks=true,linkcolor=blue!50!black,citecolor=green!40!black,urlcolor=blue!55!black]{hyperref}
\usepackage[nameinlink,capitalize,noabbrev]{cleveref}
\providecommand{\doi}[1]{doi: \href{https://doi.org/#1}{\nolinkurl{#1}}}
\newtheorem{theorem}{Theorem}[section]
\newtheorem{lemma}[theorem]{Lemma}
\newtheorem{proposition}[theorem]{Proposition}

\theoremstyle{definition}
\newtheorem{definition}[theorem]{Definition}

\makeatletter
\def\theHALG@line{\thealgorithm.\arabic{ALG@line}}
\makeatother
\theoremstyle{remark}

\hypersetup{pdftitle={Temporal Fair Division: Simultaneous Guarantees for Goods and Chores},pdfauthor={Tianqi Chen}}
\title{Temporal Fair Division:\\Simultaneous Guarantees for Goods and Chores}
\author{Tianqi Chen\thanks{School of Mathematical Science, Zhejiang University, Hangzhou 310027, P. R. China. Email: \texttt{ctq@zju.edu.cn}.}}
\date{}
\begin{document}
\maketitle
\begin{abstract}
We study temporal fair division of indivisible items in both the goods and chores settings, with \(n\ge2\) agents and nonnegative additive values or costs. Items arrive one at a time and must be assigned immediately and irrevocably. The arrival order and all agents' values or costs are known in advance, and the cumulative allocation must satisfy the fairness requirements after each arrival. For goods, we give a polynomial-time algorithm that simultaneously achieves exact temporal proportionality up to one item (TPROP1), a \(1/n\) approximation to temporal maximin share (TMMS), and a \(1/(n-1)\) approximation to temporal envy-freeness up to one item (TEF1). The TMMS factor is optimal. For chores, we give a polynomial-time algorithm that simultaneously achieves exact TPROP1 and a \(2-1/n\) approximation to TMMS. Classical scheduling lower bounds based on fixed sequences carry over to chores TMMS, establishing the optimality of this factor for \(n=2,3\). For every \(n\ge2\), we further show that \(2-1/n\) is the optimal TMMS factor subject to exact TPROP1. For every fixed \(n\ge3\) and every finite \(\beta\ge1\), we construct an instance of chores admitting no \(\beta\)-TEF1 allocation.
\end{abstract}
\section{Introduction}\label{intro:section}
Fair division studies how to allocate resources and tasks among agents with different preferences. For indivisible goods and chores, exact proportionality and envy-freeness may be impossible, motivating relaxed fairness notions such as proportionality up to one item (PROP1), envy-freeness up to one item (EF1), and maximin share (MMS).

Many allocation problems unfold over time. Goods may be delivered on different days, and tasks may become available as a project progresses. An allocation that is fair at the end may leave some agents with little value or a disproportionate workload for much of the process. This motivates requiring fairness of the cumulative allocation after every arrival. Earlier assignments may also be difficult to revise once goods have been used or tasks have been started. In planned deliveries or scheduled projects, the items and their order may be known before allocation begins.

We study the fully informed temporal model of Elkind et al.~\cite{elkind2025temporal} with one item arriving at a time. Before the first arrival, the planner knows the entire sequence and all agents' values or costs. Each item must be assigned irrevocably as soon as it arrives. We seek a single assignment of recipients for which the cumulative allocation satisfies the desired fairness guarantees after each arrival. Allocations chosen separately for successive prefixes may assign the same item to different recipients.

This need to preserve earlier decisions also arises in incremental optimization~\cite{BernsteinDGH2022,ChrobakKNY2008,LinNRW2010,MettuPlaxton2003}. For example, in the incremental median problem, all locations and distances are known in advance, but facilities are opened one at a time and cannot later be moved or closed. After each opening, the service cost should be close to the optimum for the same number of facilities. Both settings require guarantees at every stage while retaining earlier decisions.

Fairness guarantees for a single allocation need not extend to all prefixes: for goods, EF1 at every prefix need not be achievable when there are at least three agents~\cite[Theorem~4.2 and Appendix~A]{he2019fairer}. For both goods and chores, exact temporal EF1 can be guaranteed for two agents and for several restricted valuation classes~\cite{elkind2025temporal}. For general additive valuations, which exact or approximate fairness guarantees can a single allocation satisfy simultaneously at every prefix? We study temporal versions of PROP1, EF1, and MMS for both goods and chores, focusing on attainable approximation factors and polynomial-time computation.

\subsection{Three main results}
We establish three main results for temporal PROP1 (TPROP1), temporal EF1 (TEF1), and temporal MMS (TMMS), formally defined in Section~\ref{mod:section}. Larger approximation factors are better for goods, and smaller factors are better for chores.

Our first two results answer the question of Elkind et al.~\cite{elkind2025temporal} on the existence and polynomial-time computation of TPROP1 allocations for goods and chores, respectively, with additional fairness guarantees for the same allocation. Our third result resolves their TEF1 existence question for chores and rules out every finite approximation factor for each fixed \(n\ge3\).

\begin{enumerate}[leftmargin=*,itemsep=0.8em]
\item \textbf{Goods: simultaneous fairness guarantees.}
Algorithm~\ref{go:algorithm} computes in polynomial time an allocation that simultaneously satisfies exact TPROP1, \(1/n\)-TMMS, and \(1/(n-1)\)-TEF1 (Theorem~\ref{go:main}).
The TMMS factor matches the known impossibility bound~\cite{amanatidis2025personalized,ChoiLiTeh2026} and is therefore optimal. The TEF1 guarantee recovers exact TEF1 for two agents~\cite{he2019fairer} and gives \(1/2\)-TEF1 for three agents. Proposition~\ref{go:tef1-tight-example} shows that the factor \(1/(n-1)\) is tight for Algorithm~\ref{go:algorithm}.

\item \textbf{Chores: simultaneous fairness guarantees.}
Algorithm~\ref{cp:algorithm} computes in polynomial time an allocation that simultaneously satisfies exact TPROP1 and \((2-1/n)\)-TMMS (Theorem~\ref{cp:main}).
Classical scheduling lower bounds based on fixed sequences carry over to chores TMMS and establish that \(2-1/n\) is optimal for \(n=2,3\)~\cite{faigle1989partition}. We further prove that \(2-1/n\) is the optimal TMMS factor among allocations satisfying exact TPROP1 (Theorem~\ref{cp:conditional-lower}).

\item \textbf{Chores: no finite TEF1 approximation.}
For every fixed \(n\ge3\) and every finite \(\beta\ge1\), there is an instance of chores with positive integer costs that admits no \(\beta\)-TEF1 allocation (Theorem~\ref{ce:main}). For two agents, exact TEF1 allocations always exist~\cite{elkind2025temporal}.

\end{enumerate}

Exact TPROP1 is achievable for both goods and chores, but the additional guarantees differ. For goods, it can be combined with the optimal TMMS factor. For chores, our TMMS factor is optimal among exact TPROP1 allocations and is also unconditionally optimal for \(n=2,3\). The unconditional optimum for \(n\ge4\) remains open. The contrast is sharper for TEF1: for every fixed \(n\ge3\), the same goods allocation achieves a positive approximation, whereas no finite approximation can be guaranteed for chores.

Table~\ref{tab:landscape} summarizes these guarantees and impossibility results. It also includes supplementary results for temporal proportionality up to any item (TPROPX) and temporal envy-freeness up to any item (TEFX), whose definitions and proofs are given in Appendix~\ref{mod:x-variants}.

\begin{table}[htbp]
\centering
\small
\setlength{\tabcolsep}{4pt}
\renewcommand{\arraystretch}{1.16}
\begin{tabular}{@{}>{\raggedright\arraybackslash}p{0.16\linewidth}
 >{\raggedright\arraybackslash}p{\dimexpr0.42\linewidth-2\tabcolsep\relax}
 >{\raggedright\arraybackslash}p{\dimexpr0.42\linewidth-2\tabcolsep\relax}@{}}
\toprule
Fairness notion & Goods & Chores \\
\midrule
TPROP1
 & \textbf{Exact} (Thm.~\ref{go:main})
 & \textbf{Exact} (Thm.~\ref{cp:main}) \\
\addlinespace
TMMS
 & $\boldsymbol{1/n}$, tight\newline
   (Thm.~\ref{go:main}; \cite{amanatidis2025personalized,ChoiLiTeh2026})
 & $\boldsymbol{2-1/n}^{\dagger}$ (Thm.~\ref{cp:main}) \\
\addlinespace
TEF1
 & $\boldsymbol{1/(n-1)}^{*}$ (Thm.~\ref{go:main})
 & Exact for $n=2$~\cite{elkind2025temporal}\newline
   \textbf{No finite factor for $n\ge3$} (Thm.~\ref{ce:main}) \\
\addlinespace[0.7em]
TPROPX
 & No positive factor$^{\circ}$
 & $n$, tight$^{\circ}$ \\
\addlinespace
TEFX
 & No positive factor~\cite[Lemma~1]{choi2026structured}, \cite{elkind2025temporal}
 & No finite factor$^{\circ}$ \\
\bottomrule
\end{tabular}
\caption{Summary of guarantees and impossibility bounds for $n\ge2$
in the fully informed temporal model.}
\label{tab:landscape}
\vspace{3pt}
\begin{minipage}{\linewidth}
\footnotesize
Bold entries are the main results of this paper.
Algorithm~\ref{go:algorithm} simultaneously achieves the TPROP1, TMMS,
and TEF1 guarantees for goods, while Algorithm~\ref{cp:algorithm}
simultaneously achieves the TPROP1 and TMMS guarantees for chores.

\smallskip
$^{*}$ Tight for Algorithm~\ref{go:algorithm}
(Proposition~\ref{go:tef1-tight-example}).

\smallskip
$^{\dagger}$ Tight for $n=2,3$~\cite{faigle1989partition}; tight under
exact TPROP1 for every $n\ge2$ (Theorem~\ref{cp:conditional-lower}).

\smallskip
$^{\circ}$ See Appendix~\ref{mod:x-variants}.
\end{minipage}
\end{table}

\subsection{Proof overview}
Both algorithms first preprocess each agent's values or costs in arrival order and then construct a single recipient map in reverse arrival order. The preprocessing applies equal charges or service to \(n\) active items at a time. For goods, distinct representatives ensure that each agent's bundle contains a good from each of her active sets. For chores, a reverse greedy rule selects pairwise-disjoint intervals for each agent. These conditions yield lower bounds on allocated values for goods and upper bounds on allocated costs for chores.

\paragraph{Goods.} The budget process decomposes each good's value into charges and a residual budget. Using the recorded active sets and budgets, a reverse sweep maintains one distinct representative for each agent and chooses replacements with maximum available residual budget. A representative stays with its agent until its item is assigned, so each agent's final bundle intersects every active set in her budget process.

Each agent's allocated value therefore covers the cumulative common charge. At each processed prefix, at most \(n-1\) goods retain positive budgets, giving a direct bound on the MMS benchmark, while accounting for outside goods gives TPROP1. The maximum-budget rule additionally controls the residual value of a comparison bundle after removing its representative, yielding TEF1 for the same allocation.

\paragraph{Chores.} We interpret chore costs as service requirements and record each job's unfinished period as an interval. Whenever \(n\) unfinished jobs are present, the process gives them equal service until at least one finishes. The resulting intervals have overlap depth at most \(n\). A reverse greedy rule then chooses recipients while keeping the selected intervals of each agent pairwise disjoint. Its correctness follows from the depth bound, since a failed step would force \(n+1\) intervals in one row to contain a common point.

Before a selected chore arrives, all earlier selected chores have finished. Each service batch contributes to at most one of these earlier chores, although it provides equal service to \(n\) jobs. Their total cost is therefore at most \(1/n\) of the total cost before that arrival. This bound yields exact TPROP1 and the TMMS approximation guarantee.

\enlargethispage{\baselineskip}
The matching TMMS lower bound under exact TPROP1 uses identical costs. A long sequence of unit-cost chores forces every agent to receive nearly the same number of chores. One final large chore then creates a large total cost for its recipient, regardless of which agent receives it. By increasing the length of the initial sequence, the resulting cost-to-share ratio approaches \(2-1/n\).

\paragraph{Chores TEF1 impossibility.} We express costs as powers of a common base and first reason only about their integer exponents. If an agent's second-largest owned exponent exceeds the largest exponent in another agent's bundle, removing one owned chore still leaves a higher-exponent chore. Choosing the base sufficiently large turns this exponent gap into a violation of any prescribed finite TEF1 factor.

For three agents, a nine-chore instance creates this gap through three groups: preparation chores, core chores, and two final chores. For every agent, each core chore costs strictly more than every preparation chore, and each final chore costs strictly more than every core chore. The final chores constrain the ownership of the core chores, while the preparation chores make the remaining possibilities incompatible with the required exponent comparisons. For general \(n\), we combine blocks corresponding to all triples of agents and introduce anchor chores. The agents without anchors contain a triple that inherits the three-agent contradiction, while separated exponent ranges prevent other blocks from removing the violation.

\subsection{Related work}\label{intro:related}

\paragraph{Fair division of indivisible items.} Classical fair division studies the existence and computation of allocations satisfying fairness notions such as EF1, PROP1, and MMS~\cite{lipton2004approximately,budish2011combinatorial,conitzer2017fair}. Related work studies fairness and efficiency for goods, chores, and mixed manna~\cite{aziz2019fair}. Amanatidis et al.~\cite{amanatidis2023survey} provide a comprehensive survey of the literature on indivisible goods.

\paragraph{Temporal fair division.} He et al.~\cite{he2019fairer} study informed allocation and the changes needed to maintain EF1 over time, establishing goods TEF1 existence for two agents and nonexistence in general for three or more agents. Elkind et al.~\cite{elkind2025temporal} develop the temporal EF1 model for goods and chores and extend the two-agent existence result to chores. Cookson et al.~\cite{cookson2025temporal} study the compatibility of fairness within daily batches, at the final horizon, and cumulatively through each day.

Subsequent work studies temporal fairness under structural restrictions~\cite{choi2026structured}, with scheduling buffers~\cite{choi2026scheduling}, and for mixed manna~\cite{ChoiLiTeh2026}. Related models consider repeated allocations and repeated matchings of recurring items~\cite{igarashi2024repeated,caragiannis2024repeated,adams2026perpetual}. For goods TMMS, the known upper bound of \(1/n\)~\cite{amanatidis2025personalized,ChoiLiTeh2026} is attained by our algorithm together with TPROP1 and approximate TEF1.

\paragraph{Online fair division.} In our fully informed temporal model, items are assigned irrevocably as they arrive, and the planner knows all items, their order, and all agents' values or costs in advance. Fairness is evaluated on the cumulative allocation after each arrival~\cite{elkind2025temporal,he2019fairer}. Aleksandrov and Walsh~\cite{aleksandrov2020survey} survey the main online models. Work without predictions of future item values studies envy, proportionality, share guarantees, and cumulative fairness~\cite{aleksandrov2015foodbank,benade2024online,song2026onlinechores,neoh2026closing,kahana2026perpetual}. Recent work investigates high-probability or expected realized fairness against oblivious adversaries~\cite{choo2026proportionality,cohen2026oblivious,chenlong2026randomized}. Stochastic and random-order arrivals provide further settings for studying fairness and welfare~\cite{benade2024online,cohen2024randomorder}.

Another line of work provides aggregate information about future items. Known total values or costs, often expressed through normalization, have been studied through fairness guarantees and competitive analysis for indivisible allocation~\cite{zhou2023multiagent,neoh2026information,chentan2026competitive}. Maximum-item-value predictions provide another form of aggregate information for approximating proportionality~\cite{choo2026proportionality,neoh2026closing}. More detailed predictions specify each agent's multiset of item values~\cite{neoh2026information,NeohTeh2026frequency} or estimated valuation vectors~\cite{melissourgos2026predictions}. These works study how the information provided and, when predictions may be inaccurate, their errors affect achievable fairness.

\section{Model and preliminaries}\label{mod:section}
There are $n\ge2$ agents, indexed by $[n]=\{1,\ldots,n\}$. The $m\ge0$ indivisible items $g_1,\ldots,g_m$ arrive one at a time in this order. Before allocation begins, the planner knows the entire sequence and all agents' values or costs. A recipient map $a:[m]\to[n]$, where $[0]=\varnothing$, assigns $g_t$ irrevocably to agent $a(t)$ upon arrival. For $0\le t\le m$, let $G_t=\{g_1,\ldots,g_t\}$ denote the set of items in the first $t$ arrivals. At this prefix, agent $i$ holds the bundle
\[
A_i(t)=\{g_s:s\le t,\ a(s)=i\}.
\]
We seek a single map $a$ whose cumulative allocation $(A_1(t),\ldots,A_n(t))$ satisfies the required fairness inequalities for every agent at every prefix $0\le t\le m$. Throughout the paper, ties are broken in favor of the smallest index.

For goods, agent $i$ has a nonnegative additive valuation $v_i$: for a set $S$ of items, $v_i(S)=\sum_{g\in S}v_i(g)$, with each $v_i(g)$ finite and nonnegative. For chores, she has a finite nonnegative additive cost function $c_i$ with the analogous convention. Empty sets have value and cost zero. Throughout the paper the maximum of an empty set of nonnegative item values or costs is defined to be zero.

\begin{definition}[Temporal proportionality up to one item]\label{mod:tprop1}
For goods and $\alpha\in[0,1]$, an allocation is $\alpha$-TPROP1 if, for every agent $i$ and every prefix $t$,
\begin{equation}\label{mod:goods_prop}
v_i(A_i(t))+\max_{g\in G_t\setminus A_i(t)}v_i(g)
\ \ge\ \alpha\,\frac{v_i(G_t)}{n}.
\end{equation}

For chores and $\beta\ge1$, an allocation is $\beta$-TPROP1 if, for every agent $i$ and every prefix $t$,
\begin{equation*}
c_i(A_i(t))-\max_{g\in A_i(t)}c_i(g)
\ \le\ \beta\,\frac{c_i(G_t)}{n}.
\end{equation*}
The factor-one case is called exact TPROP1.
\end{definition}

\begin{definition}[Temporal envy-freeness up to one item]\label{mod:tef1}
For goods and $\alpha\in[0,1]$, an allocation is $\alpha$-TEF1 if, for every prefix $t$ and every ordered pair of distinct agents $i,j$,
\begin{equation*}
v_i(A_i(t))\ \ge\ \alpha\left(v_i(A_j(t))-
                    \max_{g\in A_j(t)}v_i(g)\right).
\end{equation*}

For chores and $\beta\ge1$, an allocation is $\beta$-TEF1 if, for every prefix $t$ and every ordered pair of distinct agents $i,j$,
\begin{equation*}
c_i(A_i(t))-\max_{g\in A_i(t)}c_i(g)
\ \le\ \beta\,c_i(A_j(t)).
\end{equation*}
The factor-one case is called exact TEF1.
\end{definition}
All one-item repairs are hypothetical and are evaluated using agent $i$'s values or costs. For goods, TPROP1 adds an outside good, whereas TEF1 removes a good from the comparison bundle; for chores, both remove a chore from the agent's own bundle. Applying the standard EF1-to-PROP1 implication at every prefix of the same
allocation shows that exact TEF1 implies exact TPROP1 for
goods~\cite[Lemma~3.6]{conitzer2017fair} and
chores~\cite[Proposition~2]{aziz2019fair}.
Under our definitions, the same averaging argument gives
$\alpha$-TEF1 $\Rightarrow$ $\alpha$-TPROP1 for goods and
$\beta$-TEF1 $\Rightarrow$ $\beta$-TPROP1 for chores.

The goods maximin-share benchmark of Budish~\citep[Definition~2]{budish2011combinatorial} is the value an agent can secure if she partitions the goods into $n$ bundles and receives a least valuable bundle. For chores, the minimax share is the smallest worst-bundle cost she can achieve by such a partition.

\begin{definition}[Temporal maximin and minimax shares]\label{mod:tmms}
Let $\Pi_n(S)$ be the set of ordered partitions $(P_1,\ldots,P_n)$ of $S$
into $n$ pairwise disjoint bundles, allowing empty bundles. For each agent
$i\in[n]$ and prefix $0\le t\le m$, define her goods maximin share and
chores minimax share, respectively, by
\begin{align}
\mu_i^g(t)&=\max_{(P_1,\ldots,P_n)\in\Pi_n(G_t)}\min_{b\in[n]}v_i(P_b),\label{mod:goods_mms}\\
\mu_i^c(t)&=\min_{(P_1,\ldots,P_n)\in\Pi_n(G_t)}\max_{b\in[n]}c_i(P_b).\notag
\end{align}
The extrema are attained because there are only finitely many partitions.
For goods and $\alpha\in[0,1]$, an allocation is $\alpha$-TMMS if, for every agent $i$ and every prefix $t$,
\[
 v_i(A_i(t))\ge\alpha\,\mu_i^g(t).
\]

For chores and $\beta\ge1$, an allocation is $\beta$-TMMS if, for every agent $i$ and every prefix $t$,
\[
 c_i(A_i(t))\le\beta\,\mu_i^c(t).
\]
The factor-one case is called exact TMMS.
\end{definition}

The following elementary properties of the MMS benchmarks will be used in the subsequent proofs.

\begin{lemma}[Elementary share bounds]\label{mod:share_bounds}
For every agent $i$, the goods benchmark satisfies $\mu_i^g(t)=0$
whenever $t<n$. At every prefix, the chores benchmark satisfies
\[
    \mu_i^c(t)\ge \frac{c_i(G_t)}{n},
    \qquad
    \mu_i^c(t)\ge \max_{g\in G_t}c_i(g).
\]
If $\mu_i^c(t)=0$, then every arrived item has zero cost to agent $i$.
\end{lemma}
\begin{proof}
When $t<n$, every partition into $n$ bundles contains an empty bundle.
Since all values are nonnegative, the minimum bundle value in every
partition is zero, giving $\mu_i^g(t)=0$.

In every partition of $G_t$, the maximum bundle cost is at least the
average $c_i(G_t)/n$. By nonnegativity, the bundle containing a chore
$g\in G_t$ has cost at least $c_i(g)$, so the maximum bundle cost is
also at least $\max_{g\in G_t}c_i(g)$.
Minimizing over all partitions gives both chores bounds.
If $\mu_i^c(t)=0$, the second bound forces every arrived cost to be zero.
\end{proof}

Standard fairness implications apply at every prefix. For goods,
$\alpha$-TEF1 implies $\frac{\alpha}{n-1+\alpha}$-TMMS~\cite[Proposition~3.6]{ijcai2018p6}.
For chores, $\beta$-TPROP1 implies $(\beta+1)$-TMMS by
Lemma~\ref{mod:share_bounds}, since
\[
c_i(A_i(t)) \le \beta\frac{c_i(G_t)}{n}+\max_{g\in A_i(t)}c_i(g) \le(\beta+1)\mu_i^c(t).
\]
Applied to our guarantees, these implications yield
$1/((n-1)^2+1)$-TMMS for goods and $2$-TMMS for chores.
The direct analyses in Sections~\ref{go:section} and~\ref{cp:section}
yield $1/n$-TMMS and $(2-1/n)$-TMMS, respectively, improving the
goods factor for $n\ge3$ and the chores factor for all $n$.

\section{Goods: simultaneous fairness guarantees}\label{go:section}

We construct one allocation satisfying exact TPROP1, \(1/n\)-TMMS, and \(1/(n-1)\)-TEF1. The preprocessing separates each good's value into charges and a residual budget. A reverse sweep then uses distinct representatives to ensure that each agent's final bundle meets every active set in her budget process. This gives the agent a share of every charge. Choosing representatives with maximum available residual budget also controls the value of a comparison bundle after deleting one good.

\subsection{Preprocessing: the scalar budget process}\label{go:scalar-section}

Assume \(m\ge n\) and fix an agent \(i\). We process her values in the original arrival order. An active good's budget is the part of its value that has not yet been charged.

\begin{enumerate}[label=\arabic*.,leftmargin=*]
\item At prefix \(n\), set \(S_i(n)=G_n\) and give each active good \(g\) budget \(v_i(g)\). Subtract the minimum budget \(\delta_i(n)\) from all \(n\) goods, and write \(b_{i,g}(n)\) for the remaining budget.
\item At each arrival \(t>n\), evict one good \(h_i(t)\in S_i(t-1)\) with zero budget, retain every other active good, and insert \(g_t\) with budget \(v_i(g_t)\). Thus
   \[
   S_i(t)=\bigl(S_i(t-1)\setminus\{h_i(t)\}\bigr)\cup\{g_t\}.
   \]
   Subtract the minimum \(\delta_i(t)\) of the resulting \(n\) budgets from all of them, and denote the remaining budgets by \(b_{i,g}(t)\).
\end{enumerate}

The set \(S_i(t)\) is the active set on which the charge at step \(t\) is applied. For each arrived good outside \(S_i(t)\), set \(b_{i,g}(t)=0\). Define the cumulative common charge by
\begin{equation}\label{go:cumulative-charge}
D_i(t)=\sum_{s=n}^{t}\delta_i(s).
\end{equation}
The following identity accounts for the value of any subset of the prefix.

\begin{lemma}[Budget accounting]\label{go:scalar}
The scalar budget process is well-defined. At every prefix \(n\le t\le m\), \(S_i(t)\) contains exactly \(n\) goods, at least one active budget is zero, and \(0\le b_{i,g}(t)\le v_i(g)\) for every \(g\in G_t\). For every \(X\subseteq G_t\),
\begin{equation}\label{go:budget-accounting}
v_i(X)
=
\sum_{s=n}^{t}\delta_i(s)|X\cap S_i(s)|
+
\sum_{g\in X}b_{i,g}(t).
\end{equation}
\end{lemma}
\begin{proof}
At initialization, subtracting the minimum of \(n\) nonnegative budgets leaves all budgets nonnegative and at least one equal to zero. At every later arrival, a zero-budget good is therefore available for eviction. Removing one good and inserting the new good preserves the active-set size, and subtracting the new minimum preserves the budget properties. A good's budget starts at its value and only decreases while it is active. Eviction sets no positive budget to zero. This proves the stated properties of the process.

For each good, every charge reduces its budget by exactly the charged amount. A good is evicted only after its budget reaches zero, so its original value equals all charges it has incurred plus its current residual budget, including after eviction. At step \(s\), the goods of \(X\) incur a total charge of \(\delta_i(s)|X\cap S_i(s)|\). Adding these charges and their residual budgets gives the identity.
\end{proof}

\paragraph{A budget example.}
Fix an agent \(i\), let \(n=3\), and consider
\[
(v_i(g_1),\ldots,v_i(g_5))=(4,1,3,2,1).
\]
Residual budgets are listed in increasing order of the goods' arrival indices.
\begin{center}
\begin{tabular}{c c c c}
\hline
\(t\) & \(S_i(t)\) & \(\delta_i(t)\) & Residual budgets\\
\hline
\(3\) & \(\{g_1,g_2,g_3\}\) & \(1\) & \((3,0,2)\)\\
\(4\) & \(\{g_1,g_3,g_4\}\) & \(2\) & \((1,0,0)\)\\
\(5\) & \(\{g_1,g_4,g_5\}\) & \(0\) & \((1,0,1)\)\\
\hline
\end{tabular}
\end{center}
At arrival four, \(g_2\) is evicted.
At arrival five, \(g_3\) is evicted under the smallest-index rule,
while \(g_4\) remains active with zero budget, so the common charge is zero.
Good \(g_1\) has incurred total charges of three and retains budget one,
accounting for its original value four.

\subsection{The allocation algorithm and its complexity}\label{go:algorithm-section}

We use the recorded active sets and budgets to choose recipients in reverse arrival order. At reverse prefix \(G_t\), agent \(i\) has a representative \(r_i(t)\in S_i(t)\), and the representatives are distinct. When the sweep reaches a representative's arrival index, the good is assigned to that agent and a replacement is selected from the preceding prefix. The remaining representatives are assigned at prefix \(n\).

\begin{algorithm}[H]
\caption{Temporal allocation of goods}
\label{go:algorithm}
\begin{algorithmic}[1]
\Require \(n\ge2\), \(m\ge0\), and the complete nonnegative value matrix \((v_i(g_t))_{i\in[n],\,t\in[m]}\).
\Ensure A recipient map \(a:[m]\to[n]\).
\If{\(m<n\)}
  \State \Return the map \(a:[m]\to[n]\) defined by \(a(t)=t\) for every \(t\in[m]\).\label{go:line:short}
\EndIf
\For{each agent \(i\in[n]\)}
  \State Run the scalar budget process of Section~\ref{go:scalar-section}; store \(S_i(t)\) and \(b_{i,g}(t)\) for \(n\le t\le m\)
\EndFor
\For{\(i=1,\ldots,n\)}
  \State Choose \(\displaystyle r_i(m)\in\operatorname*{arg\,max}_{g\in S_i(m)\setminus\{r_j(m):j<i\}}b_{i,g}(m)\)
\EndFor
\For{\(t=m,m-1,\ldots,n+1\)}\label{go:line:sweep}
  \If{\(g_t=r_i(t)\) for some agent \(i\)}
    \State \(a(t)\gets i\)
    \State Choose \(\displaystyle r_i(t-1)\in\operatorname*{arg\,max}_{g\in S_i(t-1)\setminus\{r_j(t):j\ne i\}}b_{i,g}(t-1)\)
    \State \(r_j(t-1)\gets r_j(t)\) for every \(j\ne i\)
  \Else
    \State \(a(t)\gets1\)
    \State \(r_j(t-1)\gets r_j(t)\) for every \(j\in[n]\)
  \EndIf
\EndFor\label{go:line:sweep-end}
\For{each agent \(i\in[n]\)}
  \State \(a(s)\gets i\) for the index \(s\) such that \(g_s=r_i(n)\)\label{go:line:final}
\EndFor
\State \Return \(a\)
\end{algorithmic}
\end{algorithm}

Section~\ref{go:structure-section} proves that all representative choices are available and that the representatives at prefix \(n\) exhaust \(G_n\).

\paragraph{Time and space complexity.}\label{go:implementation} For \(m\ge n\), the direct implementation uses \(O(mn^2)\) time and \(O(mn^2)\) space. For each of the \(n\) agents, a forward step updates and stores an active set of size \(n\), taking \(O(n)\) time. The complete preprocessing therefore takes \(O(mn^2)\) time and stores \(O(mn^2)\) values and identifiers. The reverse phase initializes the representatives in \(O(n^2)\) time. An array records whether each good is a current representative and, if so, which agent it represents. Each reverse step assigns one good and requires at most one scan of \(n\) candidates to select a replacement. Including the final assignments at prefix \(n\), the reverse phase takes \(O(mn)\) time. For \(m<n\), the direct assignment takes \(O(m)\) time and stores the \(O(m)\)-size recipient map.

For \(m\ge n\), the same recipient map can be computed in \(O(mn\log n)\) time and \(O(mn)\) space using min-heaps with a common offset for active budgets and constant-size reversible update records instead of full snapshots. After \(O(n^2)\) initialization, each forward row update takes \(O(\log n)\) time and stores \(O(1)\) information. Each new good takes the evicted good's slot in an array of active goods. Recording the slot, the evicted good, and the old offset lets us restore the row in \(O(1)\) time without updating the heap, since the evicted good had zero residual budget. Each reverse step restores all \(n\) rows and scans at most one of them for a replacement representative, so the sweep takes \(O(mn)\) time.

\subsection{Structural properties of the allocation}\label{go:structure-section}

We first show that the reverse sweep produces a recipient map and places an owned good in every active set. Throughout the analysis, \(A_i(t)\) denotes agent \(i\)'s bundle at prefix \(t\) under the final recipient map returned by Algorithm~\ref{go:algorithm}.

\begin{lemma}[Representatives and collected charges]\label{go:reverse}
For \(m\ge n\), Algorithm~\ref{go:algorithm} is well-defined and returns a recipient map. At every reverse prefix \(n\le t\le m\), its representatives are distinct and satisfy
\begin{equation}\label{eq:goods-representative-membership}
r_i(t)\in S_i(t)\cap A_i(t)
\qquad(i\in[n]).
\end{equation}
Moreover, for every agent \(i\) and every such prefix,
\[
v_i(A_i(t))\ge D_i(t)+b_{i,r_i(t)}(t).
\]
\end{lemma}
\begin{proof}
We first prove by descending induction on \(t=m,m-1,\ldots,n\) that the reverse construction reaches prefix \(t\) with distinct representatives \(r_i(t)\in S_i(t)\) for every agent \(i\). At \(t=m\), each active set has \(n\) goods and fewer than \(n\) representatives have already been chosen. Hence all initial choices are available and give distinct representatives in their respective active sets.

Suppose the induction claim holds at a prefix \(t>n\). The active-set update gives
\[
S_i(t)\setminus\{g_t\}\subseteq S_i(t-1)
\qquad(i\in[n]).
\]
If \(g_t\) is not a representative, all representatives remain unchanged, so they are distinct and belong to their respective active sets at prefix \(t-1\). If \(g_t=r_k(t)\) for an agent \(k\), every other representative remains unchanged and belongs to its agent's active set at prefix \(t-1\). Agent \(k\) chooses a replacement from the \(n\) goods in \(S_k(t-1)\) after excluding at most \(n-1\) other representatives. A replacement therefore exists and restores distinctness and active-set membership at prefix \(t-1\). This completes the induction.

At prefix \(n\), the \(n\) distinct representatives exhaust \(G_n\). The reverse loop in lines~\ref{go:line:sweep}--\ref{go:line:sweep-end} assigns each good with index greater than \(n\) once, and line~\ref{go:line:final} assigns each remaining good once. Thus Algorithm~\ref{go:algorithm} returns a recipient map.

Now fix an agent \(i\) and a reverse prefix \(t\), and write \(r_i(t)=g_s\). This good remains \(i\)'s representative until the sweep processes index \(s\) if \(s>n\), or reaches prefix \(n\) if \(s\le n\). In either case, Algorithm~\ref{go:algorithm} assigns \(g_s\) to \(i\). Since \(s\le t\), we have \(r_i(t)\in A_i(t)\). Together with active-set membership, this proves \eqref{eq:goods-representative-membership} for every \(n\le t\le m\).

Finally, fix an agent \(i\) and a prefix \(t\) with \(n\le t\le m\). For every \(n\le s\le t\), \eqref{eq:goods-representative-membership} and \(A_i(s)\subseteq A_i(t)\) give \(r_i(s)\in S_i(s)\cap A_i(t)\). Applying Lemma~\ref{go:scalar} to \(A_i(t)\) and using the definition of \(D_i(t)\) in~\eqref{go:cumulative-charge} gives
\[
v_i(A_i(t))=\sum_{s=n}^{t}\delta_i(s)|A_i(t)\cap S_i(s)|+\sum_{g\in A_i(t)}b_{i,g}(t)\ge D_i(t)+b_{i,r_i(t)}(t).
\]
Each intersection contains the representative \(r_i(s)\), and the residual sum contains \(b_{i,r_i(t)}(t)\).
\end{proof}

The preceding properties suffice for the TPROP1 and TMMS guarantees. To obtain TEF1 as well, we use the maximum-residual rule to control the budgets of goods outside the current representatives.

\begin{lemma}[Dominance of a maximum-residual representative]\label{go:dominance}
For every reverse prefix \(n\le t\le m\) and agent \(i\), let
\[
F_i(t)=S_i(t)\setminus\{r_j(t):j\ne i\}.
\]
Then \(r_i(t)\in F_i(t)\), and
\[
b_{i,r_i(t)}(t)\ge b_{i,g}(t)
\qquad(g\in F_i(t)).
\]
\end{lemma}
\begin{proof}
We prove the claim by descending induction on \(t=m,m-1,\ldots,n\). At prefix \(m\), all goods of \(F_i(m)\) were available when agent \(i\) selected her representative. Later agents only exclude additional goods, so her maximizing choice dominates every good remaining in \(F_i(m)\). \(r_i(m)\in F_i(m)\) follows from Lemma~\ref{go:reverse}.

Suppose the claim holds at prefix \(t>n\), and fix an agent \(i\). If \(r_i(t)=g_t\), the replacement maximizes the budget over exactly \(F_i(t-1)\), since the other representatives remain unchanged.

Otherwise, \(r_i(t-1)=r_i(t)\). Take \(g\in F_i(t-1)\). If \(g=h_i(t)\), then \(b_{i,g}(t-1)=0\), so the required inequality follows from nonnegativity. If \(g\ne h_i(t)\), the active-set update gives \(g\in S_i(t)\). We claim that \(g\in F_i(t)\). Otherwise, since \(g\in S_i(t)\), we have \(g=r_j(t)\) for some \(j\ne i\). Since \(g\in G_{t-1}\), we have \(g\ne g_t\), so this representative remains unchanged: \(r_j(t-1)=g\). This contradicts \(g\in F_i(t-1)\). Undoing the common charge and applying the induction hypothesis give
\[
b_{i,r_i(t-1)}(t-1)=b_{i,r_i(t)}(t)+\delta_i(t)\ge b_{i,g}(t)+\delta_i(t)=b_{i,g}(t-1).
\]
This proves the claim at prefix \(t-1\) in both cases.
\end{proof}

\subsection{Simultaneous guarantees and tightness}\label{go:guarantees-section}

We now derive all three guarantees for the allocation returned by Algorithm~\ref{go:algorithm}. Lemma~\ref{go:reverse} ensures that each agent collects at least one common charge at every step. The budget identity then gives TPROP1 and TMMS, while Lemma~\ref{go:dominance} supplies the additional residual-budget bound needed for TEF1.

\begin{theorem}[Simultaneous TPROP1, TMMS, and TEF1 guarantees]\label{go:main}
For every finite nonnegative additive-goods instance with \(n\ge2\) agents, Algorithm~\ref{go:algorithm} returns an allocation that is simultaneously exact TPROP1, \(1/n\)-TMMS, and \(1/(n-1)\)-TEF1.
\end{theorem}
\begin{proof}
Let \(a\) be the recipient map returned by Algorithm~\ref{go:algorithm}. Write \(M_i^{\mathrm{out}}(t)=\max_{g\in G_t\setminus A_i(t)}v_i(g)\).

At a prefix \(t<n\), the arrived goods have distinct recipients, either by line~\ref{go:line:short} or because line~\ref{go:line:final} assigns the \(n\) distinct representatives in \(G_n\) to different agents. There are at most \(n-1\) outside goods, so
\[
v_i(G_t)
\le v_i(A_i(t))+(n-1)M_i^{\mathrm{out}}(t)
\le n v_i(A_i(t))+(n-1)M_i^{\mathrm{out}}(t).
\]
This gives TPROP1 at these prefixes. Lemma~\ref{mod:share_bounds} gives \(\mu_i^g(t)=0\), so the TMMS guarantee also holds. Finally, every bundle contains at most one good, giving exact TEF1.

Now fix \(n\le t\le m\) and an agent \(i\).

For TPROP1, \eqref{eq:goods-representative-membership} and \(A_i(s)\subseteq A_i(t)\) give \(r_i(s)\in S_i(s)\cap A_i(t)\) for every \(n\le s\le t\). Thus at most \(n-1\) goods of \(S_i(s)\) lie outside \(A_i(t)\). At prefix \(t\), there are at most \(n-1\) outside active goods, each with budget at most \(M_i^{\mathrm{out}}(t)\). Applying the budget identity \eqref{go:budget-accounting} to \(G_t\setminus A_i(t)\) and using these bounds gives
\[
\begin{aligned}
v_i(G_t\setminus A_i(t))
&=\sum_{s=n}^{t}\delta_i(s)\lvert S_i(s)\setminus A_i(t)\rvert
  +\sum_{g\in G_t\setminus A_i(t)}b_{i,g}(t)\\
&\le (n-1)D_i(t)+(n-1)M_i^{\mathrm{out}}(t)\\
&\le (n-1)v_i(A_i(t))+(n-1)M_i^{\mathrm{out}}(t),
\end{aligned}
\]
where the last inequality uses Lemma~\ref{go:reverse}. Together with the short-prefix bound, this proves the strengthened proportionality inequality
\begin{equation}\label{go:strengthened-prop}
v_i(A_i(t))+\frac{n-1}{n}M_i^{\mathrm{out}}(t)
\ge \frac{v_i(G_t)}{n}
\end{equation}
at every prefix. It implies exact TPROP1.

For the TMMS guarantee, note that at most \(n-1\) arrived goods have positive residual budget at prefix \(t\), since only the \(n\) active goods can have positive budget and at least one active budget is zero. Every partition of \(G_t\) into \(n\) bundles therefore contains a bundle \(P\) with zero total residual budget. By Lemma~\ref{go:scalar},
\[
v_i(P)
=\sum_{s=n}^{t}\delta_i(s)|P\cap S_i(s)|
\le nD_i(t).
\]
The minimum bundle value in every partition is at most \(nD_i(t)\), so \(\mu_i^g(t)\le nD_i(t)\). Lemma~\ref{go:reverse} now gives
\[
v_i(A_i(t))\ge D_i(t)\ge \frac{\mu_i^g(t)}n.
\]

Finally, for the TEF1 guarantee, fix \(j\ne i\) and put \(Q=A_j(t)\setminus\{r_j(t)\}\). By \eqref{eq:goods-representative-membership}, each current representative belongs to its own agent's bundle. If \(r_k(t)\in Q\) for some \(k\), then \(k\ne j\) and \(r_k(t)\in A_k(t)\cap A_j(t)\), contradicting the disjointness of the bundles. Thus \(Q\) contains no current representative. Every good of \(Q\cap S_i(t)\) consequently belongs to \(F_i(t)\) and has budget at most \(b_{i,r_i(t)}(t)\) by Lemma~\ref{go:dominance}. There are at most \(n-1\) such goods because \(r_i(t)\in S_i(t)\setminus Q\).

By \eqref{eq:goods-representative-membership}, for every \(n\le s\le t\), the representative \(r_i(s)\) belongs to \(A_i(t)\cap S_i(s)\) and is absent from \(Q\). Thus \(|Q\cap S_i(s)|\le n-1\). Applying Lemmas~\ref{go:scalar} and~\ref{go:reverse} gives
\[
\begin{aligned}
v_i(Q)
&=\sum_{s=n}^{t}\delta_i(s)|Q\cap S_i(s)|
  +\sum_{g\in Q}b_{i,g}(t)\\
&\le (n-1)\bigl(D_i(t)+b_{i,r_i(t)}(t)\bigr)\\
&\le (n-1)v_i(A_i(t)).
\end{aligned}
\]
Deleting a largest-valued good from \(A_j(t)\) leaves value no greater than deleting \(r_j(t)\). Hence
\[
v_i(A_i(t))
\ge \frac{1}{n-1}
\left(v_i(A_j(t))-\max_{g\in A_j(t)}v_i(g)\right).
\]
Together with the short-prefix case, this establishes all three guarantees.
\end{proof}

The TPROP1 guarantee in Theorem~\ref{go:main} is exact. The \(1/n\)-TMMS factor matches the impossibility bound of Amanatidis et al.~\cite[Proposition~5.1]{amanatidis2025personalized} and is therefore optimal for \(n\ge2\). For \(n=2\), our TEF1 guarantee is exact, recovering the existence result of He et al.~\cite{he2019fairer}. For every \(n\ge3\), exact TEF1 allocations need not exist~\cite{he2019fairer}. The following proposition shows that the \(1/(n-1)\)-TEF1 guarantee is tight for Algorithm~\ref{go:algorithm}.

\begin{proposition}[Tightness for Algorithm~\ref{go:algorithm}]\label{go:tef1-tight-example}
For every \(n\ge2\), the \(1/(n-1)\)-TEF1 guarantee of Algorithm~\ref{go:algorithm} is tight.
\end{proposition}
\begin{proof}
Recall that ties are broken in favor of the smallest index. The \(2n-1\) goods arrive in the order
\[
h_1,\ldots,h_{n-2},x_1,\ldots,x_n,z.
\]
Their singleton values are given below, where \(1\le\ell\le n-2\) and \(1\le r\le n\).

\[
\begin{array}{c|ccc}
 \text{agent}&h_\ell&x_r&z\\ \hline
 1,\ldots,n-1&1&0&2\\
 n&0&1&1
\end{array}
\]

The list of \(h\) goods is empty when \(n=2\).

For each agent \(i<n\), the initial active set contains all \(h\) goods and the two zero-budget goods \(x_1,x_2\). Each subsequent \(x\) arrival evicts the earlier of the two active \(x\) goods and inserts another zero-budget good. All charges are therefore zero, and all \(h\) goods retain budget one. Just before \(z\) arrives, the two active \(x\) goods are \(x_{n-1},x_n\). The arrival of \(z\) evicts \(x_{n-1}\), while \(x_n\) keeps the minimum budget at zero. Hence
\[
S_i(m)=\{h_1,\ldots,h_{n-2},x_n,z\},\qquad
b_{i,h_\ell}(m)=1,\quad b_{i,x_n}(m)=0,\quad b_{i,z}(m)=2.
\]
At reverse initialization, agent \(1\) selects \(z\), and agents \(2,\ldots,n-1\) select \(h_1,\ldots,h_{n-2}\), respectively.

In agent \(n\)'s budget process, every \(h\) good has budget zero. The initial active set contains all \(h\) goods and \(x_1,x_2\). As \(x_3,\ldots,x_n\) arrive, the \(h\) goods are evicted in order. Each charge is zero while an \(h\) good remains. At the prefix ending with \(x_n\), which is the initialization prefix when \(n=2\), the active set consists of the \(n\) unit-valued \(x\) goods. Subtracting one leaves all their budgets zero. The arrival of \(z\) then evicts \(x_1\), and the remaining zero-budget \(x\) goods make the next charge zero. Thus
\[
S_n(m)=\{x_2,\ldots,x_n,z\},\qquad
b_{n,x_r}(m)=0\ (2\le r\le n),\quad b_{n,z}(m)=1.
\]
Since \(z\) is already selected, agent \(n\) selects \(x_2\). The initial representatives of the reverse sweep are therefore
\[
(r_1(m),\ldots,r_n(m))=(z,h_1,\ldots,h_{n-2},x_2).
\]

Every representative except agent \(1\)'s belongs to \(G_n\) and remains unchanged throughout the reverse loop in lines~\ref{go:line:sweep}--\ref{go:line:sweep-end}. Each good processed in that loop is therefore either agent \(1\)'s representative or is not a representative, and in both cases is assigned to agent \(1\). At prefix \(n\), the other representatives are the \(h\) goods and \(x_2\), so agent \(1\)'s remaining representative is \(x_1\). Line~\ref{go:line:final} gives
\[
\begin{aligned}
A_1(m)&=\{x_1,z\}\cup\{x_r:3\le r\le n\},\\
A_{k+1}(m)&=\{h_k\}\qquad(1\le k\le n-2),\\
A_n(m)&=\{x_2\}.
\end{aligned}
\]
In agent \(n\)'s valuation, her own bundle has value one, while agent \(1\)'s bundle contains \(n\) goods of value one each. Therefore
\[
\frac{v_n(A_n(m))}
{v_n(A_1(m))-\max_{g\in A_1(m)}v_n(g)}
=\frac1{n-1}.
\]
Together with Theorem~\ref{go:main}, this proves the claimed tightness.
\end{proof}

\section{Chores: simultaneous fairness guarantees}
\label{cp:section}

The second main result gives one allocation of chores that is both exact
TPROP1 and $(2-1/n)$-TMMS. Section~\ref{cp:guarantees-section} derives
these guarantees and establishes the optimality of the TMMS factor for
$n=2,3$, as well as its optimality subject to exact TPROP1 for every
$n\ge2$.
We obtain this guarantee by controlling an agent's cost just before
each chore assigned to her arrives: that cost is at most her proportional
share of the preceding prefix.  At any later prefix, deleting her last-arrived
chore leaves precisely the cost controlled by this certificate.  The same
certificate also bounds her total assigned cost relative to her minimax
share, giving both guarantees for one allocation.

For this section, write
\[
 T_i(t)=c_i(G_t)=\sum_{s=1}^{t}c_i(g_s),\qquad T_i(0)=0.
\]
The construction separates the accounting within one cost row from the
choice of recipients across rows. We index rows by agents: row $i$
represents agent $i$'s costs and the intervals constructed from them by
virtual service. We select one interval copy per chore, with
pairwise-disjoint selected intervals in each row. Selecting a chore's
interval in row $i$ assigns that chore to agent $i$.
Disjointness ensures that each service batch
is charged at most once to that agent's earlier assigned chores.
We first construct the virtual-service intervals and prove their scalar
certificate. We then present a reverse greedy allocation algorithm and
prove that it selects disjoint intervals in each row. Finally, we derive
both fairness guarantees and analyze the tightness of the TMMS factor.

\subsection{Preprocessing: virtual service and interval construction}
\label{cp:scalar-section}

The preprocessing step constructs intervals for one agent so that every
pairwise-disjoint selection satisfies the desired certificate.
Each positive batch gives equal service to exactly $n$ unfinished jobs,
including at most one selected job. Before any selected job arrives,
all earlier selected jobs have completed, so their full costs can be
accounted for using at most a $1/n$ share of the total service supplied so far.

Fix a nonnegative cost sequence $c(1),\ldots,c(m)$ for one agent.
Regard each item $t$ as a virtual job requiring $c(t)$ units of service,
and maintain an initially empty set of unfinished positive-cost jobs.
Apply steps 1 and 2 at each arrival $t=1,\ldots,m$, followed by step 3
after all arrivals have been processed.

\begin{enumerate}
    \item At arrival $t$, if $c(t)=0$, declare job $t$ complete immediately
    and set $e(t)=t$. Otherwise, insert it into the unfinished set with
    residual requirement $c(t)$.

    \item If the resulting unfinished set contains $n$ jobs, serve each
    of them by the smallest current residual requirement. For every job
    $s$ completed by this batch, set $e(s)=t$ and remove it from the set.
    No service is supplied when fewer than $n$ jobs are unfinished.

    \item After processing all arrivals, set $e(s)=m$ for every job $s$ that
    remains unfinished. Define $I(t)=[t,e(t)]$ for every $t\in[m]$.
\end{enumerate}

For $m\ge n$, from prefix $n$ onward the positive residual requirements
and positive decrement batches agree with those in the scalar budget
process of Section~\ref{go:scalar-section} on the same numerical row.
Here, completed jobs are removed immediately and their completion times
are recorded. The goods cache, however, retains zero-budget items and
evicts only one at a later arrival.

The depth of an interval family is the maximum number of intervals
containing a common point. We use closed intervals, so intervals
sharing an endpoint overlap.

\begin{lemma}[Virtual-service certificate]
\label{cp:service}
The virtual-service construction produces intervals $I(t)=[t,e(t)]$
with integer endpoints in $[m]$ and depth at most $n$.
Every pairwise-disjoint selected subfamily $\{I(s):s\in S\}$ satisfies
\begin{equation}
\label{cp:scalar}
    \sum_{\substack{s\in S\\s<q}}c(s)
    \le \frac{1}{n}\sum_{s<q}c(s)
    \qquad\text{for every }q\in S.
\end{equation}
\end{lemma}

\begin{proof}
Before each arrival there are at most $n-1$ unfinished positive-cost jobs.
This holds initially. A positive-cost arrival can produce at most $n$
such jobs. If it produces $n$, the service amount is positive and at
least one job completes, so at most $n-1$ remain afterwards.
A zero-cost arrival leaves the unfinished set unchanged.
Thus the bound is preserved, at most one service batch is needed at
each arrival, and every positive batch serves exactly $n$ jobs.

Every interval is nonempty and has integer endpoints in $[m]$.
At an arrival $u$, the only earlier intervals containing $u$ belong
to jobs unfinished just before $u$, including any jobs that complete
during the batch at $u$. There are at most $n-1$ of them.
Including the current interval gives depth at most $n$.
No interval starts strictly between arrivals, so the depth cannot
increase there.

Let $d_u$ be the service received by each participating job at arrival
$u$, with $d_u=0$ when no batch occurs.
If $d_u>0$, let $Q_u$ be the set of the $n$ jobs receiving that service;
if $d_u=0$, put $Q_u=\varnothing$.
Every interval corresponding to a job in $Q_u$ contains $u$.
Consequently, a pairwise-disjoint selected family contains at most
one of those intervals.

Fix $q\in S$. Since $I(q)$ starts at $q$, disjointness of the closed
intervals implies $e(s)<q$ for every earlier selected job $s<q$.
A job unfinished at the horizon has endpoint $m\ge q$, so no such job
is among these earlier selected jobs.
They have all completed strictly before $q$.
A positive completed job's entire cost equals its accumulated service,
and a zero-cost job contributes zero to both quantities.
We may therefore express the cost of the earlier selected jobs as a
sum over batches before $q$ and then use the fact that each batch
serves at most one selected job:
\[
    \sum_{\substack{s\in S\\s<q}}c(s)
    =\sum_{u<q}d_u\bigl|Q_u\cap\{s\in S:s<q\}\bigr|
    \le\sum_{u<q}d_u.
\]
To bound this sum of batch amounts, count the service given to all jobs.
All service before $q$ is given to jobs that arrived before $q$,
and no job receives more than its required service.
Since each positive batch serves exactly $n$ jobs,
\[
    n\sum_{u<q}d_u\le\sum_{s<q}c(s).
\]
Combining the above two bounds proves
\eqref{cp:scalar}.
\end{proof}

\paragraph{A service example.}
Let $n=3$ and $(c(1),\ldots,c(5))=(4,1,3,2,1)$.
There are exactly two positive batches:
\begin{center}
\begin{tabular}{c c c c l}
\hline
$u$ & $Q_u$ & $d_u$ & completed jobs & residuals after the batch\\
\hline
$3$ & $\{1,2,3\}$ & $1$ & $\{2\}$ & job 1: 3, job 3: 2\\
$4$ & $\{1,3,4\}$ & $2$ & $\{3,4\}$ & job 1: 1\\
\hline
\end{tabular}
\end{center}
After arrival five, jobs 1 and 5 remain unfinished.
The intervals, in order, are
\[
    [1,5],\quad [2,3],\quad [3,4],\quad [4,4],\quad [5,5].
\]
For the disjoint selection $S=\{2,4,5\}$ and $q=5$, the earlier
selected jobs 2, 4 have completed, and their total cost is
$1+2=3\le10/3$.
Each positive batch serves exactly one of them.
Job 5 need not have completed: its own cost is excluded from the
left-hand side of \eqref{cp:scalar}.
The intervals of jobs 3, 4 meet at endpoint four, so they cannot
both be selected.
Allowing that pair would require, at $q=4$,
$c(3)=3\le(4+1+3)/3=8/3$, which is false.
This illustrates why intervals are closed and earlier selected jobs
must satisfy $e(s)<q$.

\subsection{The allocation algorithm and its complexity}
\label{cp:algorithm-section}

Applying the virtual-service construction of Section~\ref{cp:scalar-section}
to each agent's cost row gives intervals $I_i(t)=[t,e_i(t)]$.
By Lemma~\ref{cp:service}, the intervals in each row have depth at most $n$,
and every pairwise-disjoint selection in that row satisfies the
service certificate.
It remains to choose one interval copy for each chore so that the
selected intervals in every row are pairwise disjoint.
Choosing $I_i(t)$ assigns chore $t$ to agent $i$.

We construct the recipient map using a reverse greedy algorithm and
prove its feasibility in Section~\ref{cp:selection-section}. Before
processing index $t$, only the chores with indices greater than $t$
have been assigned. For each agent $i$, let $\tau_i$ be the smallest
arrival index among these chores assigned to her, with $\tau_i=+\infty$
if there are none. The interval of chore $t$ is disjoint from all intervals
already selected for $i$ exactly when
\[
 e_i(t)<\tau_i.
\]
Among the agents satisfying this condition, the rule chooses one with the
smallest $\tau_i$.

\begin{algorithm}[H]
\caption{Temporal allocation of chores}
\label{cp:algorithm}
\begin{algorithmic}[1]
\Require $n\ge2$, $m\ge0$, and the complete nonnegative cost matrix $(c_i(g_t))_{i\in[n],\,t\in[m]}$.
\Ensure A recipient map $a:[m]\to[n]$.
\If{$m=0$}
  \State \Return the empty map
\EndIf
\For{each agent $i\in[n]$}
  \State Run the virtual-service construction of Section~\ref{cp:scalar-section} on row $i$; store all $e_i(t)$\label{cp:line:preprocess}
\EndFor
\State $\tau_i\gets+\infty$ for every agent $i\in[n]$
\For{$t=m,m-1,\ldots,1$}
  \State Choose $\displaystyle i\in\operatorname*{arg\,min}_{j\in[n]:\,e_j(t)<\tau_j}\tau_j$
  \State $a(t)\gets i$
  \State $\tau_i\gets t$
\EndFor
\State \Return $a$
\end{algorithmic}
\end{algorithm}

\paragraph{Time and space complexity.}
For $m\ge1$, the direct implementation uses $O(mn^2)$ time and $O(mn)$
space. In the preprocessing of line~\ref{cp:line:preprocess}, each arrival inserts at most one job and triggers at
most one service batch. Scanning and updating at most $n$ unfinished jobs
takes $O(n)$ time per arrival in each row, for $O(mn^2)$ time over all
rows. The reverse sweep scans the $n$ agents once per chore to find an
eligible agent with minimum $\tau_i$, and therefore takes $O(mn)$ time.
Storing the $mn$ endpoints, the recipient map, and the $n$ current values
$\tau_i$ uses $O(mn)$ space. The unfinished jobs can be maintained one row
at a time in $O(n)$ additional space.

The same recipient map can be computed in $O(mn\log n)$ time and $O(mn)$
space by implementing virtual service with a min-heap. For each row,
maintain a cumulative service clock and store each positive-cost job at
a completion coordinate equal to the clock at its insertion plus its
cost. When $n$ jobs are unfinished, advance the clock to the smallest
active coordinate and remove every job at that coordinate. Each job is
inserted and removed at most once, and the heap contains at most $n$ jobs.
This gives $O(m\log n)$ time per row and produces the same endpoints;
the reverse sweep remains unchanged.

\subsection{Correctness of interval selection}
\label{cp:selection-section}

The eligibility condition preserves pairwise disjointness among each
agent's selected intervals. It remains to show that an eligible agent
exists at every reverse step. A failed step would force $n+1$ intervals
in one row to contain a common point, contradicting the depth bound
from Lemma~\ref{cp:service}.

\begin{lemma}[Feasibility of Algorithm~\ref{cp:algorithm}]\label{cp:interval}
Algorithm~\ref{cp:algorithm} is well-defined and returns a recipient map
$a:[m]\to[n]$. For every agent $i$, the intervals $I_i(t)$ with $a(t)=i$
are pairwise disjoint.
\end{lemma}

\begin{proof}
By Lemma~\ref{cp:service}, the interval family produced for each agent has
depth at most $n$. At each successful reverse step, the chosen interval
ends before the smallest arrival index already assigned to its recipient,
so it is disjoint from every interval previously selected for that agent.

Suppose the reverse sweep reaches an index $t$ with no eligible agent.
At this step,
\[
 e_i(t)\ge\tau_i\qquad(i\in[n]).
\]
All $\tau_i$ are finite, since an agent with $\tau_i=+\infty$ would be
eligible. These indices are distinct because each processed job has one
recipient. Choose $r$ with the largest $\tau_r$, and put $q=\tau_r$.
Then $a(q)=r$, no index in $\{t+1,\ldots,q-1\}$ has been assigned to $r$,
and $t<\tau_i<q$ for every $i\ne r$.

For each $i\ne r$, define
\[
 u_i=\max\{s:t<s<q,\ a(s)=i\}.
\]
This set is nonempty because it contains $\tau_i$. Thus $u_i$ is the
largest arrival index below $q$ assigned to agent $i$ in the already
constructed suffix.

When the reverse sweep processed $u_i$, agent $r$'s value of $\tau_r$
was $q$. Agent $i$'s value of $\tau_i$ was strictly greater than $q$,
possibly $+\infty$, because no index strictly between $u_i$ and $q$
was assigned to $i$, and $q$ was assigned to $r$. If $e_r(u_i)<q$, then
$r$ would have been eligible with a smaller value of $\tau_r$ than the
value of $\tau_i$. The greedy rule could not then have chosen $i$.
Consequently,
\[
 e_r(u_i)\ge q\qquad(i\ne r).
\]

The intervals $I_r(u_i)$ all contain $q$, since $u_i<q$. The failed step
gives $e_r(t)\ge q$, so $I_r(t)$ also contains $q$, and $I_r(q)$ contains
its starting point $q$. The indices $u_i$ are distinct because they
have different recipients, and all lie strictly between $t$ and $q$.
Hence $I_r(t)$, $I_r(q)$, and $I_r(u_i)$ for
$i\in[n]\setminus\{r\}$ are $n+1$ distinct intervals in row $r$
containing $q$, a contradiction.

An eligible agent therefore exists at every reverse step, and
Algorithm~\ref{cp:algorithm} assigns every chore while preserving pairwise
disjointness for each agent.
\end{proof}

\subsection{Simultaneous guarantees and tightness}
\label{cp:guarantees-section}

The disjoint intervals selected by Algorithm~\ref{cp:algorithm} bound each
agent's assigned cost immediately before every chore she receives. We use
this bound to derive TPROP1 and TMMS for the same allocation, and then
examine the tightness of the TMMS factor.

\begin{theorem}[Simultaneous TPROP1 and TMMS guarantees]\label{cp:main}
For every finite nonnegative additive-chores instance with $n\ge2$
agents, Algorithm~\ref{cp:algorithm} returns an allocation that is
simultaneously exact TPROP1 and $(2-1/n)$-TMMS.
\end{theorem}

\begin{proof}
Let $a$ be the recipient map returned by Algorithm~\ref{cp:algorithm}.
The case $m=0$ is immediate, since all bundles are empty and all prefix
costs are zero. Lemma~\ref{cp:interval} guarantees that the intervals
selected for each agent are pairwise disjoint. For any arrival $q$
assigned to agent $i$, the certificate of Lemma~\ref{cp:service} gives
\begin{equation}\label{cp:strong}
 \sum_{\substack{s<q\\a(s)=i}}c_i(g_s)
 \leq \frac{T_i(q-1)}{n}.
\end{equation}

To obtain TPROP1 from~\eqref{cp:strong}, fix an agent $i$ and a prefix $t$.
If $A_i(t)$ is nonempty, let $q$ be the arrival index of its last chore.
There are no further chores assigned to $i$ between $q$ and $t$. Thus
\begin{align*}
 c_i(A_i(t))-\max_{g\in A_i(t)}c_i(g)
 &\leq c_i(A_i(t)\setminus\{g_q\})\\
 &=\sum_{\substack{s<q\\a(s)=i}}c_i(g_s)\\
 &\leq\frac{T_i(q-1)}{n}
 \leq\frac{T_i(t)}{n},
\end{align*}
where the first inequality uses $\max_{g\in A_i(t)}c_i(g)\geq c_i(g_q)$
and the last uses nonnegative costs.
If $A_i(t)$ is empty, its residual cost is zero.  These arguments apply
to every agent and every prefix of the same allocation, proving exact
TPROP1.

For TMMS, we bound each agent's full bundle cost using the same certificate.
The common share bounds proved in Lemma~\ref{mod:share_bounds},
with $T_i(t)=c_i(G_t)$, give
\begin{equation}\label{cp:mms-lower}
 \mu_i^c(t)\geq\frac{T_i(t)}n,
 \qquad
 \mu_i^c(t)\geq c_i(g_s)\quad(s\leq t).
\end{equation}
If $A_i(t)$ is nonempty, let $q$ again be the index of its last chore
and put $c=c_i(g_q)$. Certificate~\eqref{cp:strong} bounds the cost assigned
before $q$.  Moreover, $G_t$ contains $g_q$ as well as all of $G_{q-1}$,
so nonnegativity gives $T_i(q-1)\leq T_i(t)-c$.  Hence
\begin{align*}
 c_i(A_i(t))
 &\leq\frac{T_i(q-1)}n+c\\
 &\leq\frac{T_i(t)-c}{n}+c\\
 &=\frac{T_i(t)}n+\left(1-\frac1n\right)c\\
 &\leq\mu_i^c(t)+\left(1-\frac1n\right)\mu_i^c(t)\\
 &=\left(2-\frac1n\right)\mu_i^c(t).
\end{align*}
The fourth line uses~\eqref{cp:mms-lower} and $1-1/n\geq0$.
An empty assigned bundle has cost zero.  If the minimax share is zero,
\eqref{cp:mms-lower} implies that every arrived cost in that row is
zero, so the inequality also holds in that case. Thus the same allocation
is also $(2-1/n)$-TMMS.
\end{proof}

\paragraph{Scheduling lower bounds.}
With identical cost functions, chores TMMS is equivalent to incremental
makespan minimization on $n$ identical machines with the entire job
sequence known in advance.
The classical fixed sequences $(1,1,2)$ and $(1,1,1,3,3,3,6)$ rule out
TMMS factors below $3/2$ and $5/3$ for two and three agents,
respectively~\cite[Section~6]{faigle1989partition}. Thus the factor in
Theorem~\ref{cp:main} is optimal for $n=2,3$.
The fixed-sequence constructions of Rudin and
Chandrasekaran~\cite{rudin2003improved} and Rudin~\cite{rudin2001improved}
further yield unconditional lower bounds of $\sqrt{3}$ for $n=4$ and
$1.88$ as $n\to\infty$, respectively; their applicability to the
incremental model is discussed by
Janke~\cite[Sections~1.1 and~1.4]{janke2026lower}.
Janke's four-machine lower bounds above $\sqrt{3}$~\cite{janke2026lower}
rely on branching adversaries that adapt future job sizes to earlier
assignments, so they do not directly transfer to the incremental model
or to our fully informed temporal model.

The next theorem shows that $2-1/n$ is the best TMMS factor compatible
with exact TPROP1 for every $n\ge2$. Its construction uses identical
costs and compares only the prefixes immediately before and after the
final chore arrives.

\begin{theorem}[TMMS lower bound under exact TPROP1]
\label{cp:conditional-lower}
For every fixed integer $n\ge2$ and every real
$1\le\beta<2-1/n$, there is a fixed sequence of chores with identical
strictly positive additive costs such that every allocation
satisfying exact TPROP1 at all prefixes violates $\beta$-TMMS at the
terminal prefix.
\end{theorem}
\begin{proof}
Fix $n\ge2$ and $\beta<2-1/n$, and put
\[
 \Delta=2-\frac1n-\beta>0.
\]
Choose a positive integer $K$ with $K>1/(n\Delta)$. Define
\[
 L=(n-1)K,\qquad T=nL-1=n(n-1)K-1.
\]
The first $T$ chores each have cost $1$ to every agent. The final
chore $g_{T+1}$ has cost $nK$ to every agent. Thus the instance
has $T+1=n(n-1)K$ chores. All singleton costs are positive integers, and $T\ge1$.

Consider any allocation that is exact TPROP1 at every prefix.
At prefix $T$, let $x_i=|A_i(T)|$ be the number of unit chores assigned
to agent $i$. If $x_i>0$, deleting a most costly chore removes exactly
one unit. Exact TPROP1 at this prefix therefore gives
\begin{align*}
 x_i-1&\le\frac{T}{n}=L-\frac1n.
\end{align*}
Since $x_i$ is an integer, $x_i\le L$. The same bound holds when
$x_i=0$. Summing over the other $n-1$ agents now yields, for every $i$,
\begin{equation}\label{cp:forced-load}
 x_i=T-\sum_{j\ne i}x_j
 \ge nL-1-(n-1)L=L-1.
\end{equation}

At the terminal prefix, every agent's minimax share is exactly $nK$.
Indeed, the final chore alone has cost $nK$, giving the lower bound
$\mu_i^c(T+1)\ge nK$ by Lemma~\ref{mod:share_bounds}. For the reverse
bound, form a partition into $n$ bundles as follows. One bundle
contains only the final chore. Of the remaining $n-1$ bundles,
$n-2$ contain $nK$ unit chores each and one contains $nK-1$ unit
chores. These counts sum to
\[
 (n-2)nK+(nK-1)=n(n-1)K-1=T,
\]
so the partition uses every unit chore exactly once and has maximum
cost $nK$. When $n=2$, the group of $n-2$ bundles is empty and the
remaining bundle contains $2K-1$ unit chores; the same argument
applies. Thus
\begin{equation}\label{cp:terminal-share}
 \mu_i^c(T+1)=nK\qquad(i\in[n]).
\end{equation}

Let $r$ be the recipient of the final chore. Irrevocability preserves
the unit chores already assigned to $r$. By~\eqref{cp:forced-load},
her terminal cost is at least
\[
 c_r(A_r(T+1))\ge L-1+nK=(2n-1)K-1.
\]
Combining this inequality with~\eqref{cp:terminal-share} gives
\begin{align*}
 \frac{c_r(A_r(T+1))}{\mu_r^c(T+1)}
 &\ge\frac{(2n-1)K-1}{nK}\\
 &=2-\frac1n-\frac{1}{nK}\\
 &>2-\frac1n-\Delta=\beta.
\end{align*}
The strict inequality follows from the chosen $K>1/(n\Delta)$.
Hence this allocation violates $\beta$-TMMS at the terminal prefix.
\end{proof}

\section{Chores: no finite TEF1 approximation}\label{ce:section}
We next consider whether the simultaneous TPROP1 and TMMS guarantees
for chores can be extended to include approximate TEF1.
For every fixed number of agents at
least three and every prescribed finite factor $\beta\ge1$, we construct
a fixed instance with no $\beta$-TEF1 allocation. We first establish an
exponent comparison, then construct a nine-chore obstruction for three
agents and extend it to every $n\ge3$.

Define the residual cost of a bundle $B$ by
\[
 \rho_i(B)=c_i(B)-\max_{g\in B}c_i(g).
\]

\begin{theorem}\label{ce:main}
For every integer $n\ge3$ and every finite real $\beta\ge1$ there is a
fixed instance of $m=n-3+9\binom n3$ ordered chores with positive integer
additive costs such that every allocation has some prefix $t$ and
ordered pair $i\ne j$ satisfying
\[
 \rho_i(A_i(t))>\beta c_i(A_j(t)).
\]
Thus this instance admits no $\beta$-TEF1 allocation.
\end{theorem}

\subsection{Exponent comparisons}\label{ce:exponent-section}
The next lemma reduces the cost comparison to an ordering of exponents.
The relevant exponent in an agent's own bundle is the second largest:
at least one chore with that exponent or a larger one survives every
single-chore deletion.

\begin{lemma}\label{ce:exponentgap}
Suppose an instance has $m$ chores and all costs are
$c_i(g)=R^{\gamma_i(g)}$, where $R>m\beta$ and every exponent is a nonnegative
integer. At a prefix, let $A_i$ and $A_j$ be the bundles of two distinct
agents $i$ and $j$, with $|A_i|\ge2$. Let $s$ be the second-largest exponent
in $A_i$, and let $f$ be the largest exponent in $A_j$, both evaluated in
row $i$, with $f=-\infty$ if $A_j=\varnothing$.
If $s>f$, the $\beta$-EF1 inequality from $i$ to $j$ fails at that prefix.
\end{lemma}
\begin{proof}
There are two owned chores with exponents at least $s$. Deleting any one
owned chore leaves at least one of those two, so the minimum residual
cost is at least $R^s>0$. If $j$ owns no chore, this is a strict
violation against zero. Otherwise $j$ owns at most $m$ chores, each of
cost at most $R^f$, and hence $c_i(A_j)\le m R^f$. Since $s$ and $f$ are
integers, $s>f$ gives $s\ge f+1$. Also $R>m\beta\ge1$. Therefore
\[
 \rho_i(A_i)\ge R^s\ge R^{f+1}
       >\beta m R^f\ge\beta c_i(A_j).\qedhere
\]
\end{proof}

Call an allocation \emph{ordinally admissible} if, at every prefix,
the second-largest owned exponent of every agent with at least two
chores is at most every opposing bundle's largest exponent in that
agent's row. The lemma implies that every $\beta$-TEF1 allocation is
ordinally admissible whenever $R>m\beta$. This is a necessary condition
for $\beta$-TEF1; the proof will only use the implication from an exponent
gap to a violation of the cost inequality.

Equivalently, for every row $i$ and integer threshold $v$, if $i$ has
received two chores of exponent at least $v$, then every other agent must
already have received a chore of exponent at least $v$ in row $i$.
Indeed, a failed threshold gives $s\ge v>f$. Conversely, a gap $s>f$
violates that threshold condition at $v=s$. This threshold formulation
gives an equivalent way to read the exponent comparisons below.

\subsection{The three-agent case}\label{ce:base-section}
The three-agent obstruction uses the following exponent matrix, with
columns in arrival order:
\begin{equation}\label{ce:base-matrix}
 \Gamma=(\gamma_{kh})=\begin{pmatrix}
 0&1&1&1&4&3&2&6&6\\
 1&0&3&2&5&4&5&6&6\\
 2&0&1&2&3&4&4&6&6
 \end{pmatrix}.
\end{equation}
Chores $1,\ldots,4$ prepare the bundles, chores $5,6,7$ form the
core, and chores $8,9$ both have exponent six in every row. For each
agent, every core chore costs strictly more than every preparation chore,
and each of the last two chores costs strictly more than every core chore.

\Needspace{7\baselineskip}
\begin{lemma}\label{ce:base}
For the exponent matrix~\eqref{ce:base-matrix}, no recipient map
$a:[9]\to[3]$ is ordinally admissible.
Consequently, for three agents, costs $R^{\gamma_{ih}}$ with any integer
$R>9\beta$ admit no $\beta$-TEF1 allocation.
\end{lemma}
\begin{proof}
Suppose an ordinally admissible allocation exists.
We first determine the ownership of the three core chores, using the
last two chores, and then rule out every possible preparation prefix.
All comparisons below use the observer's own row.

\paragraph{Each agent receives one core chore.}
In each row, every core exponent is strictly greater than every
exponent among chores $1,\ldots,4$.
If an agent received at least two core chores, then at prefix seven
her second-largest owned exponent would be at least the smaller
of those two core exponents. With three core chores and three agents,
some other agent would have no core chore. That agent's bundle would
contain only preparation chores, or be empty. Its maximum in the
observer's row would therefore be strictly smaller.
This contradicts ordinal admissibility. Hence each agent receives
exactly one of chores $5,6,7$.

Chores $8,9$ must have different recipients. If one agent received
both, her second-largest owned exponent would be six at prefix nine.
Every other agent would hold only chores $1,\ldots,7$, all of whose
exponents are at most five in every row. This is again a strict gap.

\paragraph{A cycle forces the core ownership.}
Each agent has a unique least costly core chore in her row:
\[
 q_1=7,\qquad q_2=6,\qquad q_3=5.
\]
Map each agent $i$ to the owner of $q_i$.
Because the $q_i$ are distinct and each agent owns one core chore,
this map is a permutation of the agents.
Suppose it has a nontrivial cycle. Every agent $i$ on this cycle owns
a core chore other than $q_i$. The cycle contains at least two
agents, leaving at most one agent outside it. Since chores $8,9$
have distinct recipients, at least one is assigned to an agent on
the cycle. Let $i$ be the first cycle agent to receive one of them,
and let $j$ be the owner of $q_i$, the next agent on the cycle.
At that prefix, $j$ has not received either of chores $8,9$.

The exponent of agent $i$'s own core chore is therefore strictly larger
than that of $q_i$. All preparation exponents in row $i$ are smaller than
every core exponent, while the newly received chore has exponent six.
Thus $i$'s second-largest owned exponent is her own core exponent.
Agent $j$ owns exactly the core chore $q_i$, possibly along with
preparation chores, and has no exponent-six chore. Her bundle's
maximum in row $i$ is therefore the exponent of $q_i$.
The resulting strict gap is impossible.

The permutation consequently has no nontrivial cycle and must be
the identity. Core ownership is forced to be
\begin{equation*}
 a(5)=3,\qquad a(6)=2,\qquad a(7)=1.
\end{equation*}

\paragraph{Only two preparation cases remain.}
The first three recipients are distinct. Otherwise, at the first
repetition one agent would own two chores while another still
owned none, giving a finite second-largest exponent against
$-\infty$.

Agent $3$ cannot receive chore $1$ or chore $4$.
Indeed, in row $3$ either has exponent two, and her forced core
chore $5$ has exponent three. If she owned either preparation chore,
then at prefix five her second-largest exponent would be at least two.
At least one of the other two agents would own neither chore $1$
nor chore $4$: at most one of these two chores could remain outside
agent $3$'s bundle. That other agent would have only chores $2,3$,
or be empty, with row-$3$ maximum at most one.
This would be a strict gap. Thus agent $3$ initially receives
either chore $2$ or chore $3$, and $a(4)\ne3$.

\emph{Case 1: agent $3$ receives chore $3$.}
At prefix five, agent $3$ owns chores $3,5$, whose row-$3$
exponents are one and three. Her second-largest exponent is one.
The agent initially receiving chore $2$ has exponent zero in row $3$.
To avoid a strict gap against that agent, chore $4$ must also be
given to her, since chore $5$ goes to agent $3$.
If this agent were agent $1$, she would own chores $2,4$ at
prefix four, both of exponent one in row $1$; agent $2$ would
own only chore $1$, of row-$1$ exponent zero.
That would already violate ordinal admissibility.
Hence the first four recipients must be $1,2,3,2$.

Now give chore $6$ to its forced recipient, agent $2$.
At prefix six, agent $2$ owns chores $2,4,6$, with row-$2$
exponents $0,2,4$. Her second-largest exponent is two.
Agent $1$ still owns only chore $1$, of row-$2$ exponent one.
This strict gap rules out Case 1.

\emph{Case 2: agent $3$ receives chore $2$.}
Agent $2$ cannot initially receive chore $3$. If she did, then after
her forced receipt of chore $6$ her second-largest exponent would
be at least three, from row-$2$ exponents three and four.
Agent $1$ would own chore $1$ and possibly chore $4$, whose
row-$2$ exponents are one and two. Chores $5,6$ belong to agents
$3,2$, respectively. The maximum of agent $1$'s bundle in row $2$
would therefore be at most two, a contradiction.
The first three recipients are consequently $2,3,1$.

We already know that $a(4)\ne3$. If chore $4$ goes to agent $1$,
then she owns chores $3,4$, both with row-$1$ exponent one,
while agent $2$ owns only chore $1$, of row-$1$ exponent zero.
If chore $4$ goes to agent $2$, then she owns chores $1,4$,
with row-$2$ exponents one and two, while agent $3$ owns only
chore $2$, of row-$2$ exponent zero.
Either choice gives a strict gap at prefix four.
This rules out Case 2.

The two cases exhaust the possible initial chore of agent $3$.
Thus no ordinally admissible allocation exists. Applying
Lemma~\ref{ce:exponentgap} with $m=9$ proves the cost impossibility.
\end{proof}

\subsection{Extension to \texorpdfstring{$n$}{n} agents}\label{ce:extension-section}
We extend the obstruction in Section~\ref{ce:base-section} to every
$n\ge3$, using the exponent comparison of
Section~\ref{ce:exponent-section}, to prove Theorem~\ref{ce:main}.
The construction first presents
$n-3$ high-cost \emph{anchors}. In any ordinally admissible allocation,
these anchors must have distinct owners, leaving a triple of agents
without anchors. The construction then presents a block for every triple
of agents. The triple without anchors depends on the allocation, but its
block is already present in the fixed input. The exponent scales will
force that block to remain within the triple and make its local
violation persist after all preceding chores are included.

\begin{proof}[Proof of Theorem~\ref{ce:main}]
Let $N=\binom n3$, and fix the lexicographic enumeration $T_0,\ldots,T_{N-1}$ of all three-element
subsets of $[n]$. Use increasing agent order for local labels $1,2,3$
within each triple. The number of chores is $m=n-3+9N$. Choose
\[
 R=\lfloor m\beta\rfloor+1>m\beta.
\]
The instance starts with $n-3$ anchors, followed by one nine-chore block
for each triple in the fixed enumeration. Every anchor has exponent
$9N+1$ in every row. In block $b$, item $h\in\{1,\ldots,9\}$ has exponent
\begin{equation*}
 \gamma_i(b,h)=
 \begin{cases}
  9b+1+\gamma_{kh},&\text{if $i\in T_b$ has local label $k$},\\
  9b+h,&\text{if $i\notin T_b$},
 \end{cases}
\end{equation*}
where $\gamma_{kh}$ is defined in~\eqref{ce:base-matrix}.
Its cost in row $i$ is $R^{\gamma_i(b,h)}$. Every exponent is a nonnegative
integer, so all costs are positive integers.

\paragraph{Exponent separation.}
The base exponents range from zero to six. Thus, in block $b$, the
exponents in the rows of agents in $T_b$ lie between $9b+1$ and $9b+7$,
and those in the rows of agents outside $T_b$ lie between $9b+1$ and
$9b+9$. In every row, a nonanchor item from an earlier
block therefore has exponent at most $9b$: for a block $b'<b$, its
exponent is at most $9b'+9\le9b$. Over all blocks the nonanchor exponents
are at most $9N$, strictly less than the anchor exponent $9N+1$.
Finally, in the row of an agent outside $T_b$, the current exponent $9b+h$
strictly exceeds every preceding nonanchor exponent in that row.
Earlier blocks have exponents at most $9b$, and preceding items in the
same block have exponents $9b+1,\ldots,9b+h-1$.

Assume, for a contradiction, that a recipient map $a$ is $\beta$-TEF1.
By Lemma~\ref{ce:exponentgap}, $a$ is ordinally admissible.
We first identify the triple without anchors, then force all nine chores
of its block into that triple, and finally apply the base obstruction to
the complete bundles at the resulting prefix.

\paragraph{The anchors leave exactly three agents anchorless.}
All $n-3$ anchors must have different owners. If an anchor were the
second received by one agent, that agent would have two owned chores
while some other agent remained empty, since fewer than $n$ chores have
arrived during the anchor stage. The second-largest owned exponent
would then exceed the opposing maximum $-\infty$, contradicting
ordinal admissibility. Precisely $n-3$ agents therefore own one anchor
each, and the remaining three agents own none. Denote those three by
$T^*$. If $n=3$, the anchor stage is empty and $T^*=[3]$; all subsequent
arguments retain their meaning.

\paragraph{Every item in the $T^*$ block stays inside $T^*$.}
The fixed enumeration contains $T^*$; let $b$ be its index. Suppose
item $h$ of this block is assigned to an agent $\ell\notin T^*$. This
agent already owns an anchor of exponent $9N+1$, strictly larger than
the current exponent $9b+h$ in row $\ell$. Thus agent $\ell$'s second-largest
owned exponent is at least $9b+h$, regardless of the other chores she owns.

Choose any $j\in T^*$. This agent has no anchor and does not receive
the current item. All chores in $j$'s bundle precede the current one and
are nonanchors. Because $\ell\notin T^*$, row $\ell$ uses strictly increasing
exponents throughout this block. The separation bounds therefore place
all those row-$\ell$ exponents strictly below $9b+h$. If $j$'s bundle is empty, its maximum is $-\infty$, with the same
strict comparison. Thus agent $\ell$'s second-largest owned exponent
exceeds the row-$\ell$ maximum of $j$'s bundle, a contradiction.
The comparison uses only the absence of anchors in $T^*$ and the strict
increase of the exponents in row $\ell$. It therefore applies to each item
regardless of how preceding items were assigned. All nine items must
therefore be assigned inside $T^*$.

\paragraph{The base contradiction survives all preceding history.}
Restricting the recipient map to these nine chores and relabeling the
agents by their local labels gives a recipient map from $[9]$ to $[3]$.
Lemma~\ref{ce:base} supplies a block prefix and a pair $i\ne j$ in $T^*$ for which $i$ has at least two block items,
with base second-largest exponent $s$ strictly greater than the
row-$i$ base maximum $f$ in $j$'s block bundle. The value $f=-\infty$
is permitted. Compare their complete bundles at this global prefix.

The two largest block exponents owned by $i$ are both at least
$9b+1+s$. Adding earlier items cannot remove those two items, so the
second-largest exponent in $i$'s complete bundle is at least $9b+1+s$.
Neither $i$ nor $j$ owns an anchor, and every earlier-block exponent in
row $i$ is at most $9b$.

If $j$ has no current-block item at the witness prefix, its entire
bundle has maximum at most $9b$, or $-\infty$ if it is empty. Since
$s\ge0$, the second-largest exponent of $i$ is at least $9b+1$ and
is strictly larger than that maximum.

If $j$ has a current-block item, its largest such exponent in row $i$
is exactly $9b+1+f$. Here $f\ge0$, so this exceeds all earlier-block
exponents. It is therefore also the maximum of $j$'s complete bundle
in row $i$. The strict inequality $s>f$ shows that $i$'s second-largest
owned exponent again exceeds it.

Both cases contradict ordinal admissibility and hence $\beta$-TEF1.
This proves Theorem~\ref{ce:main}. The number of chores is exactly
$n-3+9\binom n3$ as claimed.
\end{proof}

\section{Conclusion}\label{intro:discussion}
The three main results describe which fairness requirements can coexist across all prefixes of one allocation. For goods, a maximum-residual reverse construction simultaneously gives exact TPROP1, the optimal TMMS factor $1/n$, and $1/(n-1)$-TEF1. For chores, reverse greedy interval selection and a service certificate give exact TPROP1 and TMMS factor $2-1/n$ in the same allocation. This factor is optimal subject to exact TPROP1 for every $n\ge2$, and is also optimal without that requirement for $n=2,3$. Both positive guarantees are achieved by polynomial-time algorithms. In contrast, for every fixed $n\ge3$, no finite uniform TEF1 factor exists for chores, even without imposing any other fairness condition.

The optimality statements have different scopes. For goods, our $1/n$-TMMS guarantee is optimal~\cite{amanatidis2025personalized,ChoiLiTeh2026}. The $1/(n-1)$-TEF1 guarantee is tight for Algorithm~\ref{go:algorithm} by Proposition~\ref{go:tef1-tight-example}, but determining the optimal TEF1 factor for $n\ge3$ remains open.
For chores, Theorem~\ref{cp:conditional-lower} shows that the $2-1/n$ TMMS factor is optimal subject to exact TPROP1. Classical scheduling constructions give unconditional TMMS lower bounds of $3/2$ for $n=2$ and $5/3$ for $n=3$~\cite{faigle1989partition}, as well as $\sqrt3$ for $n=4$ and $1.88$ as $n\to\infty$~\cite{rudin2003improved,rudin2001improved}. Together with Theorem~\ref{cp:main}, these bounds determine the optimal unconditional TMMS factors for $n=2,3$, while the optimum for $n\ge4$ remains open.

Another open problem is whether general additive mixed-manna instances
always admit one allocation that is exact TPROP1 at every prefix in our
fully informed temporal model, where the same item may be a good for
some agents and a chore for others. At each prefix, an agent's utility
must reach $1/n$ of her utility for all items in that prefix after at most one
hypothetical correction: adding a positively valued outside item or
removing a negatively valued owned item. Temporal mixed manna has been
studied for TEF1 under structural restrictions~\cite{ChoiLiTeh2026}, and
the two-agent case admits TEF1~\cite[Theorem~D.2]{elkind2025temporal} and hence TPROP1.
If such TPROP1 allocations always exist for $n\ge3$, can they be
computed in polynomial time?

\bibliographystyle{plainnat}
\setlength{\bibsep}{6pt plus 2pt minus 1pt}
\begingroup
\interlinepenalty=10000
\bibliography{references}

@inproceedings{aziz2019fair,
 author = {Aziz, Haris and Caragiannis, Ioannis and Igarashi, Ayumi and Walsh, Toby},
 title = {Fair Allocation of Indivisible Goods and Chores},
 booktitle = {Proceedings of the 28th International Joint Conference on Artificial Intelligence},
 pages = {53--59},
 year = {2019},
 doi = {10.24963/ijcai.2019/8}
}

@article{budish2011combinatorial,
 author = {Budish, Eric},
 title = {The Combinatorial Assignment Problem: Approximate Competitive Equilibrium from Equal Incomes},
 journal = {Journal of Political Economy},
 volume = {119},
 number = {6},
 pages = {1061--1103},
 year = {2011},
 doi = {10.1086/664613},
 url = {https://ericbudish.org/files/combinatorial_assignment_problem.pdf}
}

@inproceedings{elkind2025temporal,
 author = {Elkind, Edith and Lam, Alexander and Latifian, Mohamad and Neoh, Tzeh Yuan and Teh, Nicholas},
 title = {Temporal Fair Division of Indivisible Items},
 booktitle = {Proceedings of the 24th International Conference on Autonomous Agents and Multiagent Systems},
 year = {2025},
 pages = {676--685},
 note = {Full version: \href{https://arxiv.org/abs/2410.14593v1}{arXiv:2410.14593}},
 url = {https://www.ifaamas.org/Proceedings/aamas2025/pdfs/p676.pdf}
}

@misc{ChoiLiTeh2026,
 author = {Choi, Kui-Wang and Li, Minming and Teh, Nicholas},
 title = {Temporal Fair Division of Indivisible Mixed Manna: Tractable Settings},
 year = {2026},
 eprint = {2608.20033},
 archivePrefix = {arXiv},
 primaryClass = {cs.GT},
 howpublished = {\href{https://arxiv.org/abs/2608.20033v1}{arXiv:2608.20033}}
}

@inproceedings{he2019fairer,
 author = {He, Jiafan and Procaccia, Ariel D. and Psomas, Alexandros and Zeng, David},
 title = {Achieving a Fairer Future by Changing the Past},
 booktitle = {Proceedings of the 28th International Joint Conference on Artificial Intelligence},
 year = {2019},
 pages = {343--349},
 doi = {10.24963/ijcai.2019/49},
 note = {Full version: \url{https://www.cs.cmu.edu/~cpsomas/adjustments.pdf}}
}

@article{cookson2025temporal,
 author = {Cookson, Benjamin and Ebadian, Soroush and Shah, Nisarg},
 title = {Temporal Fair Division},
 journal = {Proceedings of the AAAI Conference on Artificial Intelligence},
 volume = {39},
 number = {13},
 pages = {13727--13734},
 year = {2025},
 doi = {10.1609/aaai.v39i13.33500},
 note = {Full version: \href{https://arxiv.org/abs/2410.23416v1}{arXiv:2410.23416}}
}

@inproceedings{choi2026structured,
 author = {Choi, Kui-Wang and Li, Minming},
 title = {Temporal Fair Division of Indivisible Goods with Structured Constraints},
 booktitle = {Algorithmic Game Theory (SAGT 2026)},
 series = {Lecture Notes in Computer Science},
 volume = {16911},
 pages = {95--112},
 publisher = {Springer},
 year = {2027},
 note = {Full version: \href{https://arxiv.org/abs/2607.17224v1}{arXiv:2607.17224}},
 doi = {10.1007/978-3-032-38782-0_6}
}

@inproceedings{amanatidis2025personalized,
 author = {Amanatidis, Georgios and Lolos, Alexandros and Markakis, Evangelos and Turmel, Victor},
 title = {Online Fair Division for Personalized 2-Value Instances},
 booktitle = {Algorithmic Game Theory (SAGT 2025)},
 series = {Lecture Notes in Computer Science},
 volume = {15953},
 pages = {209--227},
 publisher = {Springer},
 year = {2026},
 doi = {10.1007/978-3-032-03639-1_12},
 note = {Full version: \href{https://arxiv.org/abs/2505.22174v1}{arXiv:2505.22174}}
}

@misc{choi2026scheduling,
 author = {Choi, Kui-Wang and Li, Minming},
 title = {Temporal Fair Division of Indivisible Goods with Scheduling},
 year = {2026},
 eprint = {2601.12835},
 archivePrefix = {arXiv},
 primaryClass = {cs.GT},
 howpublished = {\href{https://arxiv.org/abs/2601.12835v3}{arXiv:2601.12835}}
}

@inproceedings{conitzer2017fair,
 author = {Conitzer, Vincent and Freeman, Rupert and Shah, Nisarg},
 title = {Fair Public Decision Making},
 booktitle = {Proceedings of the 2017 ACM Conference on Economics and Computation},
 pages = {629--646},
 year = {2017},
 doi = {10.1145/3033274.3085125},
 url = {https://www.rupertfreeman.com/fair_public_decisions_EC17.pdf}
}

@inproceedings{adams2026perpetual,
 author = {Adams, Terrence and Segal-Halevi, Erel},
 title = {Perpetually Fair Assignments Via Balanced Sequences of Permutations},
 booktitle = {Algorithmic Game Theory (SAGT 2026)},
 series = {Lecture Notes in Computer Science},
 volume = {16911},
 pages = {1--18},
 publisher = {Springer},
 year = {2027},
 doi = {10.1007/978-3-032-38782-0_1},
 note = {Full version: \href{https://arxiv.org/abs/2602.21687v3}{arXiv:2602.21687}}
}

@misc{kahana2026perpetual,
 author = {Kahana, Ido and Segal-Halevi, Erel and Hazon, Noam},
 title = {Perpetual Fully-Online Approximate Fairness},
 year = {2026},
 eprint = {2605.19844},
 archivePrefix = {arXiv},
 primaryClass = {cs.GT},
 howpublished = {\href{https://arxiv.org/abs/2605.19844v5}{arXiv:2605.19844}}
}

@misc{NeohTeh2026frequency,
 author = {Neoh, Tzeh Yuan and Teh, Nicholas},
 title = {Online Fair Division with Frequency Predictions},
 year = {2026},
 eprint = {2610.34625},
 archivePrefix = {arXiv},
 primaryClass = {cs.GT},
 howpublished = {\href{https://arxiv.org/abs/2610.34625v1}{arXiv:2610.34625}}
}

@article{MettuPlaxton2003,
  author = {Mettu, Ramgopal R. and Plaxton, C. Greg},
  title = {The Online Median Problem},
  journal = {SIAM Journal on Computing},
  volume = {32},
  number = {3},
  pages = {816--832},
  year = {2003},
  doi = {10.1137/S0097539701383443}
}

@article{ChrobakKNY2008,
  author = {Chrobak, Marek and Kenyon, Claire and Noga, John and Young, Neal E.},
  title = {Incremental Medians via Online Bidding},
  journal = {Algorithmica},
  volume = {50},
  number = {4},
  pages = {455--478},
  year = {2008},
  doi = {10.1007/s00453-007-9005-x}
}

@article{LinNRW2010,
  author = {Lin, Guolong and Nagarajan, Chandrashekhar and Rajaraman, Rajmohan and Williamson, David P.},
  title = {A General Approach for Incremental Approximation and Hierarchical Clustering},
  journal = {SIAM Journal on Computing},
  volume = {39},
  number = {8},
  pages = {3633--3669},
  year = {2010},
  doi = {10.1137/070698257}
}

@article{BernsteinDGH2022,
  author = {Bernstein, Aaron and Disser, Yann and Gro{\ss}, Martin and Himburg, Sandra},
  title = {General Bounds for Incremental Maximization},
  journal = {Mathematical Programming},
  volume = {191},
  number = {2},
  pages = {953--979},
  year = {2022},
  doi = {10.1007/s10107-020-01576-0}
}

@article{faigle1989partition,
  author = {Faigle, Ulrich and Kern, Walter and Tur{\'a}n, Gy{\"o}rgy},
  title = {On the performance of on-line algorithms for partition problems},
  journal = {Acta Cybernetica},
  volume = {9},
  number = {2},
  pages = {107--119},
  year = {1989},
  url = {https://cyber.bibl.u-szeged.hu/index.php/actcybern/article/view/3359}
}

@article{rudin2003improved,
  author = {Rudin, III, John F. and Chandrasekaran, R.},
  title = {Improved Bounds for the Online Scheduling Problem},
  journal = {SIAM Journal on Computing},
  volume = {32},
  number = {3},
  pages = {717--735},
  year = {2003},
  doi = {10.1137/S0097539702403438}
}

@phdthesis{rudin2001improved,
  author = {Rudin, III, John F.},
  title = {Improved bounds for the on-line scheduling problem},
  school = {The University of Texas at Dallas},
  year = {2001}
}

@misc{janke2026lower,
  author = {Janke, Maximilian},
  title = {Lower Bounds Above $\sqrt{3}$ for Online Makespan Minimization on Four Identical Machines},
  year = {2026},
  eprint = {2611.10440},
  archivePrefix = {arXiv},
  primaryClass = {cs.DS},
  howpublished = {\href{https://arxiv.org/abs/2611.10440v1}{arXiv:2611.10440}}
}

@inproceedings{lipton2004approximately,
  author = {Richard J. Lipton and Evangelos Markakis and Elchanan Mossel and Amin Saberi},
  title = {On Approximately Fair Allocations of Indivisible Goods},
  booktitle = {Proceedings of the 5th ACM Conference on Electronic Commerce},
  pages = {125--131},
  year = {2004},
  doi = {10.1145/988772.988792},
  url = {https://www.cs.cmu.edu/~arielpro/mfai_papers/LMMS03.pdf}
}

@article{amanatidis2023survey,
  author = {Georgios Amanatidis and Haris Aziz and Georgios Birmpas and Aris Filos-Ratsikas and Bo Li and Herv{\'e} Moulin and Alexandros A. Voudouris and Xiaowei Wu},
  title = {Fair Division of Indivisible Goods: Recent Progress and Open Questions},
  journal = {Artificial Intelligence},
  volume = {322},
  pages = {103965},
  year = {2023},
  doi = {10.1016/j.artint.2023.103965}
}

@article{igarashi2024repeated,
  author = {Ayumi Igarashi and Martin Lackner and Oliviero Nardi and Arianna Novaro},
  title = {Repeated Fair Allocation of Indivisible Items},
  journal = {Proceedings of the AAAI Conference on Artificial Intelligence},
  volume = {38},
  number = {9},
  pages = {9781--9789},
  year = {2024},
  doi = {10.1609/aaai.v38i9.28837}
}

@article{caragiannis2024repeated,
  author = {Ioannis Caragiannis and Shivika Narang},
  title = {Repeatedly Matching Items to Agents Fairly and Efficiently},
  journal = {Theoretical Computer Science},
  volume = {981},
  pages = {114246},
  year = {2024},
  doi = {10.1016/j.tcs.2023.114246}
}

@article{aleksandrov2020survey,
  author = {Martin Aleksandrov and Toby Walsh},
  title = {Online Fair Division: A Survey},
  journal = {Proceedings of the AAAI Conference on Artificial Intelligence},
  volume = {34},
  number = {9},
  pages = {13557--13562},
  year = {2020},
  doi = {10.1609/aaai.v34i09.7081}
}

@inproceedings{aleksandrov2015foodbank,
  author = {Martin Aleksandrov and Haris Aziz and Serge Gaspers and Toby Walsh},
  title = {Online Fair Division: Analysing a Food Bank Problem},
  booktitle = {Proceedings of the 24th International Joint Conference on Artificial Intelligence},
  pages = {2540--2545},
  year = {2015},
  url = {https://ijcai.org/proceedings/2015/360}
}

@article{benade2024online,
  author = {Gerdus Benad{\`e} and Aleksandr M. Kazachkov and Ariel D. Procaccia and Alexandros Psomas and David Zeng},
  title = {Fair and Efficient Online Allocations},
  journal = {Operations Research},
  volume = {72},
  number = {4},
  pages = {1438--1452},
  year = {2024},
  doi = {10.1287/opre.2022.0332}
}

@inproceedings{song2026onlinechores,
  author = {Jiaxin Song and Biaoshuai Tao and Wenqian Wang and Yuhao Zhang},
  title = {Online {MMS} Allocation for Chores},
  booktitle = {Proceedings of the 27th ACM Conference on Economics and Computation},
  year = {2026},
  note = {Full version: \href{https://arxiv.org/abs/2507.14039}{arXiv:2507.14039}}
}

@misc{neoh2026closing,
  author = {Tzeh Yuan Neoh and Nicholas Teh},
  title = {Closing Gaps in Online Fair Division},
  year = {2026},
  eprint = {2609.05310},
  archivePrefix = {arXiv},
  primaryClass = {cs.GT},
  howpublished = {\href{https://arxiv.org/abs/2609.05310v2}{arXiv:2609.05310}}
}

@inproceedings{choo2026proportionality,
  author = {Davin Choo and Winston Fu and Tzeh Yuan Neoh and Tze-Yang Poon and Nicholas Teh},
  title = {Approximate Proportionality in Online Fair Division},
  booktitle = {Proceedings of the 43rd International Conference on Machine Learning},
  series = {Proceedings of Machine Learning Research},
  volume = {306},
  pages = {20417--20450},
  publisher = {PMLR},
  year = {2026},
  url = {https://proceedings.mlr.press/v306/choo26a.html}
}

@misc{cohen2026oblivious,
  author = {Saar Cohen and Nicholas Teh and Michael Wooldridge},
  title = {Online Fair Division under an Oblivious Adversary},
  year = {2026},
  eprint = {2609.28333},
  archivePrefix = {arXiv},
  primaryClass = {cs.GT},
  howpublished = {\href{https://arxiv.org/abs/2609.28333v2}{arXiv:2609.28333}}
}

@misc{chenlong2026randomized,
  author = {Tianqi Chen and Jingxiao Long},
  title = {Randomized Online Fair Division: High-Probability and Expected Realized Fairness},
  year = {2026},
  eprint = {2609.23577},
  archivePrefix = {arXiv},
  primaryClass = {cs.GT},
  howpublished = {\href{https://arxiv.org/abs/2609.23577v2}{arXiv:2609.23577}}
}

@inproceedings{cohen2024randomorder,
  author = {Saar Cohen and Noa Agmon},
  title = {Near-Optimal Online Resource Allocation in the Random-Order Model: Extended Abstract},
  booktitle = {Proceedings of the 23rd International Conference on Autonomous Agents and Multiagent Systems},
  pages = {2219--2221},
  year = {2024},
  url = {https://www.ifaamas.org/Proceedings/aamas2024/pdfs/p2219.pdf}
}

@inproceedings{zhou2023multiagent,
  author = {Shengwei Zhou and Rufan Bai and Xiaowei Wu},
  title = {Multi-agent Online Scheduling: {MMS} Allocations for Indivisible Items},
  booktitle = {Proceedings of the 40th International Conference on Machine Learning},
  series = {Proceedings of Machine Learning Research},
  volume = {202},
  pages = {42506--42516},
  publisher = {PMLR},
  year = {2023},
  url = {https://proceedings.mlr.press/v202/zhou23a.html}
}

@inproceedings{neoh2026information,
  author = {Tzeh Yuan Neoh and Jannik Peters and Nicholas Teh},
  title = {Online Fair Division with Additional Information},
  booktitle = {Proceedings of the 43rd International Conference on Machine Learning},
  series = {Proceedings of Machine Learning Research},
  volume = {306},
  pages = {92207--92241},
  publisher = {PMLR},
  year = {2026},
  url = {https://proceedings.mlr.press/v306/neoh26a.html}
}

@misc{chentan2026competitive,
  author = {Tianqi Chen and Zhiyi Tan},
  title = {Competitive Analysis for Online Fair Division under Multiple Fairness Notions},
  year = {2026},
  eprint = {2606.15404},
  archivePrefix = {arXiv},
  primaryClass = {cs.GT},
  howpublished = {\href{https://arxiv.org/abs/2606.15404v1}{arXiv:2606.15404}}
}

@inproceedings{melissourgos2026predictions,
  author = {Themistoklis Melissourgos and Nicos Protopapas},
  title = {Online {EFX} Allocations with Predictions: Extended Abstract},
  booktitle = {Proceedings of the 25th International Conference on Autonomous Agents and Multiagent Systems},
  pages = {3519--3521},
  year = {2026},
  doi = {10.65109/PGMH4700},
  url = {https://www.ifaamas.org/Proceedings/aamas2026/pdfs/PGMH4700.pdf},
  note = {Full version: \href{https://arxiv.org/abs/2508.04779}{arXiv:2508.04779}}
}

@inproceedings{ijcai2018p6,
  title     = {Comparing Approximate Relaxations of Envy-Freeness},
  author    = {Georgios Amanatidis and Georgios Birmpas and Vangelis Markakis},
  booktitle = {Proceedings of the Twenty-Seventh International Joint Conference on Artificial Intelligence, {IJCAI-18}},
  publisher = {International Joint Conferences on Artificial Intelligence Organization},
  pages     = {42--48},
  year      = {2018},
  doi       = {10.24963/ijcai.2018/6}
}

@article{caragiannis2019unreasonable,
  author  = {Ioannis Caragiannis and David Kurokawa and Herv{\'e} Moulin and Ariel D. Procaccia and Nisarg Shah and Junxing Wang},
  title   = {The Unreasonable Fairness of Maximum {Nash} Welfare},
  journal = {ACM Transactions on Economics and Computation},
  volume  = {7},
  number  = {3},
  pages   = {12:1--12:32},
  year    = {2019},
  doi     = {10.1145/3355902}
}

@article{aziz2024almost,
  author  = {Haris Aziz and Bo Li and Herv{\'e} Moulin and Xiaowei Wu and Xinran Zhu},
  title   = {Almost proportional allocations of indivisible chores: Computation, approximation and efficiency},
  journal = {Artificial Intelligence},
  volume  = {331},
  pages   = {104118},
  year    = {2024},
  doi     = {10.1016/j.artint.2024.104118}
}
\endgroup
\clearpage
\appendix
\section{Supplementary bounds for TPROPX and TEFX}
\label{mod:x-variants}
The following definitions and short arguments supply the TPROPX and
TEFX entries of Table~\ref{tab:landscape}. They are separate from the
three main proof chains.

\begin{definition}[TPROPX and TEFX]
\label{mod:x-definition}
For goods and $\alpha\in[0,1]$, an allocation is $\alpha$-TPROPX if, for every
agent $i$, prefix $t$, and good $g\in G_t\setminus A_i(t)$,
\[
 v_i(A_i(t))+v_i(g)\ge\alpha\,\frac{v_i(G_t)}n.
\]
If the outside set is empty, the added value is taken to be zero;
the resulting inequality holds automatically because
$A_i(t)=G_t$. For chores and $\beta\ge1$, an allocation is $\beta$-TPROPX
if, for every agent $i$, prefix $t$, and chore $g\in A_i(t)$,
\[
 c_i(A_i(t))-c_i(g)\le\beta\,\frac{c_i(G_t)}n.
\]
An empty own bundle satisfies this condition.

For goods and $\alpha\in[0,1]$, an allocation is $\alpha$-TEFX if, for every
prefix $t$, distinct agents $i,j$, and good $g\in A_j(t)$,
\[
 v_i(A_i(t))\ge\alpha\,v_i(A_j(t)\setminus\{g\}).
\]
For chores and $\beta\ge1$, an allocation is $\beta$-TEFX if, for every such
$i,j,t$ and chore $g\in A_i(t)$,
\[
 c_i(A_i(t)\setminus\{g\})\le\beta\,c_i(A_j(t)).
\]
When the bundle from which an item is to be deleted is empty, the
comparison holds automatically. The factor-one cases are called exact
TPROPX and exact TEFX, respectively.
\end{definition}
The repair must now work for every candidate item, rather than for at
least one as in TPROP1 and TEF1. We include zero-valued or zero-cost items.
Some definitions instead consider only items of positive value or cost
to the agent whose inequality is being tested
\cite{caragiannis2019unreasonable,aziz2024almost}.
Under additivity, this amounts to testing the candidate with the smallest
positive value or cost, adding it for goods TPROPX and removing it otherwise.
The bounds below hold under either convention, since all impossibility
instances are strictly positive and every allocation of chores is
$n$-TPROPX.

\begin{proposition}[TPROPX impossibility bounds]\label{obs:tpropx}
For every fixed $n\ge2$, the following statements hold.
\begin{enumerate}[label=(\roman*),leftmargin=*]
\item For every $\alpha\in(0,1]$, there is a fixed instance of goods with
identical, strictly positive additive valuations and no $\alpha$-TPROPX allocation.
Thus no positive uniform goods factor is possible.
\item Every allocation of chores is $n$-TPROPX. For every
$1\le\beta<n$, there is a fixed instance of chores with identical,
strictly positive additive costs and no $\beta$-TPROPX allocation.
Thus the optimal chores TPROPX factor is $n$.
\end{enumerate}
\end{proposition}
\begin{proof}
For goods, fix $\alpha>0$ and choose constants
\[
 H>\frac n\alpha,\qquad L>\frac{n(1+H)}\alpha.
\]
Let three goods of values $1,H,L$ arrive in that order, identically
valued by every agent. At prefix two,
$\alpha(1+H)/n>1$. If the first two goods have the same recipient,
an empty agent fails the TPROPX comparison when the hypothetical
outside good is the unit good. Hence the first two recipients must
be distinct. For $n\ge3$, an agent is still empty at this prefix
and the same unit-good comparison fails, proving the claim already.
For $n=2$, the recipients are distinct. After the last good arrives,
let $i$ be the agent who does not receive it. Her own good and the
earlier good held by the other agent have combined value $1+H$.
The latter is an eligible outside good, but
\[
 1+H<\frac{\alpha L}{n}
     <\alpha\,\frac{1+H+L}{n}.
\]
The required comparison therefore fails.

For chores, nonnegativity gives, for any allocation and any own chore $g$,
\[
 c_i(A_i(t))-c_i(g)\le c_i(G_t)
        =n\,\frac{c_i(G_t)}n.
\]
This proves the factor $n$, including empty bundles and zero-total rows.
To rule out a smaller factor, fix $1\le\beta<n$ and choose an integer
\[
 R>\frac{n}{n-\beta}>1.
\]
Take $n+1$ chores with identical costs $c_i(g_t)=R^{t-1}$ for
$t=1,\ldots,n+1$. At every prefix,
\[
 \frac{\beta}{n}\sum_{s=1}^{t}R^{s-1}
 <\frac{\beta R}{n(R-1)}R^{t-1}
 <R^{t-1}.
\]
Every allocation gives at least two of the $n+1$ chores to the same agent. At an
arrival $t$ whose recipient $i$ already owns an earlier chore $h$,
deleting $h$ leaves the new chore, so
\[
 c_i(A_i(t))-c_i(h)\ge R^{t-1}
       >\beta\,\frac{c_i(G_t)}n.
\]
This contradicts $\beta$-TPROPX.
\end{proof}

\begin{proposition}[TEFX impossibility]
\label{obs:tefx}
For every fixed $n\ge2$, no positive uniform goods TEFX factor and
no finite uniform chores TEFX factor are possible, even for identical
strictly positive valuations or costs.
\end{proposition}
\begin{proof}
Use $n$ unit items followed by one item of size $H$, with identical
values or costs. Suppose first that an allocation satisfies $\alpha$-TEFX for
goods, where $\alpha>0$. At prefix $n$, any repeated recipient would
leave an empty agent: deleting one unit good from the repeated
recipient's bundle leaves value at least one, contradicting the empty
agent's TEFX inequality. Thus each agent holds one unit good. Choose
$H>1/\alpha$. At the next prefix, an agent other than the recipient
of the last good has value one. Deleting the recipient's unit good
leaves value $H$, and $1<\alpha H$, a contradiction.

For chores, fix a finite $\beta\ge1$ and choose $H>\beta$. Again, at
prefix $n$, a repeated recipient would hold at least two unit chores
and some agent would be empty. Deleting any one of those chores leaves
cost at least one, exceeding $\beta$ times the empty bundle's cost zero.
Hence every agent holds one unit chore. The recipient of the last
chore then retains cost $H>\beta$ after her unit chore is deleted,
whereas any other agent has cost one. This violates $\beta$-TEFX.
\end{proof}

\end{document}